\documentclass[iop,revtex4]{emulateapj}
\usepackage{graphicx,color}
\usepackage{amssymb}
\usepackage{amsmath}
\usepackage{threeparttable}
\usepackage{lscape}
\usepackage{longtable}
\usepackage{changepage}
\usepackage{hyperref}
\hypersetup{colorlinks   = true, citecolor    = blue}

\newcommand{\para}{1.0}
\renewcommand{\baselinestretch}{\para}
\mathchardef\mhyphen="2D

\newcommand{\oiii}{[O\,{\sc iii}]}
\newcommand{\ovi}{O\,{\sc vi}}
\newcommand{\heii}{He\,{\sc ii}}

\newcommand{\siv}{S\,{\sc iv}}
\newcommand{\siiv}{Si\,{\sc iv}}

\newcommand{\pv}{P\,{\sc v}}

\newcommand{\ciii}{C\,{\sc iii}}
\newcommand{\civ}{C\,{\sc iv}}
\newcommand{\mgii}{Mg\,{\sc ii}}

\newcommand{\angstrom}{\text{ \normalfont\AA}}

\mathchardef\mhyphen="2D

\definecolor{blk}{rgb}{0.0,0.0,0.0}
\definecolor{red}{rgb}{0.75,0.0,0.0}
\definecolor{yel}{rgb}{0.65,0.65,0.0}
\definecolor{grn}{rgb}{0.0,0.75,0.0}
\definecolor{blu}{rgb}{0.0,0.0,0.75}
\definecolor{gry}{rgb}{0.75,0.75,0.75}

\def\lya{Ly$\alpha$}
\def\ly{$\lambda$}
\def\hi{H\,{\sc i}}

\def\heii{He\,{\sc ii}}

\def\ciii{C\,{\sc iii}}
\def\civ{C\,{\sc iv}}
\def\niii{N\,{\sc iii}}
\def\niv{N\,{\sc iv}}
\def\nv{N\,{\sc v}}
\def\oi{O\,{\sc i}}
\def\oiii{O\,{\sc iii}}
\def\oiv{O\,{\sc iv}}
\def\ov{O\,{\sc v}}
\def\ovi{O\,{\sc vi}}

\def\neiv{Ne\,{\sc iv}}
\def\nev{Ne\,{\sc v}}
\def\nevi{Ne\,{\sc vi}}
\def\neviii{Ne\,{\sc viii}}

\def\naix{Na\,{\sc ix}}

\def\mgx{Mg\,{\sc x}}
\def\mgii{Mg\,{\sc ii}}

\def\alxi{Al\,{\sc xi}}
\def\siiv{Si\,{\sc iv}}
\def\Siii{Si\,{\sc ii}}

\def\sixii{Si\,{\sc xii}}
\def\pv{P\,{\sc v}}
\def\siv{S\,{\sc iv}}
\def\sv{S\,{\sc v}}
\def\svi{S\,{\sc vi}}
\def\ariv{Ar\,{\sc iv}}
\def\arv{Ar\,{\sc v}}
\def\arvi{Ar\,{\sc vi}}
\def\arvii{Ar\,{\sc vii}}
\def\arviii{Ar\,{\sc viii}}
\def\caiv{Ca\,{\sc iv}}
\def\cav{Ca\,{\sc v}}
\def\cavi{Ca\,{\sc vi}}
\def\cavii{Ca\,{\sc vii}}
\def\caviii{Ca\,{\sc viii}}

\def\cax{Ca\,{\sc x}}

\def\nh{\ifmmode n_\mathrm{\scriptscriptstyle H} \else $n_\mathrm{\scriptscriptstyle H}$\fi}%this \nh is better, h is smaller.
\def\ne{\ifmmode n_\mathrm{\scriptstyle e} \else $n_\mathrm{\scriptstyle e}$\fi}
\def\Qh{\ifmmode Q_\mathrm{\scriptstyle H} \else $Q_\mathrm{\scriptstyle H}$\fi}
\def\Uh{\ifmmode U_\mathrm{\scriptstyle H} \else $U_\mathrm{\scriptstyle H}$\fi}
\def\Nh{\ifmmode N_\mathrm{\scriptstyle H} \else $N_\mathrm{\scriptstyle H}$\fi}
\def\Uhhp{\ifmmode U_\mathrm{\scriptstyle H,HP} \else $U_\mathrm{\scriptstyle H,HP}$\fi}
\def\Nhhp{\ifmmode N_\mathrm{\scriptstyle H,HP} \else $N_\mathrm{\scriptstyle H,HP}$\fi}
\def\Uhvhp{\ifmmode U_\mathrm{\scriptstyle H,VHP} \else $U_\mathrm{\scriptstyle H,VHP}$\fi}
\def\Nhvhp{\ifmmode N_\mathrm{\scriptstyle H,VHP} \else $N_\mathrm{\scriptstyle H,VHP}$\fi}
\def\Nion{\ifmmode N_\mathrm{\scriptstyle ion} \else $N_\mathrm{\scriptstyle ion}$\fi}

\def\Zsun{\ifmmode {\rm Z}_{\odot} \else Z$_{\odot}$\fi}
\def\Msun{\ifmmode {\rm M}_{\odot} \else M$_{\odot}$\fi}
\def\kms{\ifmmode {\rm km~s}^{-1} \else km~s$^{-1}$\fi}
\def\Lya{\ifmmode {\rm Ly}\alpha \else Ly$\alpha$\fi}
\def\Lyb{\ifmmode {\rm Ly}\beta \else Ly$\beta$\fi}
\def\Lyg{\ifmmode {\rm Ly}\gamma \else Ly$\gamma$\fi}
\def\Lyd{\ifmmode {\rm Ly}\delta \else Ly$\delta$\fi}
\def\neaod{\ifmmode n_\mathrm{\scriptscriptstyle AOD} \else $n_\mathrm{\scriptscriptstyle AOD}$\fi}
\def\necrit{\ifmmode n_\mathrm{\scriptstyle cr} \else $n_\mathrm{\scriptstyle cr}$\fi}
\def\ncr{\ifmmode n_\mathrm{\scriptstyle cr} \else $n_\mathrm{\scriptstyle cr}$\fi}
\def\nepi{\ifmmode n_\mathrm{\scriptscriptstyle PI} \else $n_\mathrm{\scriptscriptstyle PI}$\fi}
\def\gtorder{\mathrel{\raise.3ex\hbox{$>$}\mkern-14mu\lower0.6ex\hbox{$\sim$}}}
\def\ltorder{\mathrel{\raise.3ex\hbox{$<$}\mkern-14mu\lower0.6ex\hbox{$\sim$}}}
\newcommand{\vy}[2]{#1_{\scriptscriptstyle #2}}%see email small subscript

\newcommand{\SSS} {SSS}
\newcommand{\XUV} {EUV500}
\newcommand{\comp}{system}
\newcommand{\Comp}{System}
\newcommand{\comps}{systems}
\newcommand{\Comps}{Systems}
\newcommand{\sub}[2]{\ifmmode #1_\mathrm{\scriptstyle #2} \else $#1_\mathrm{\scriptstyle #2}$\fi}

\slugcomment{Submitted to ApJS 2019 Jul 14; Accepted 2019 Sep 24}

\shorttitle{ }
\shortauthors{Xu et al.}
\shortauthors{}

\begin{document}


\title{HST/COS observations of quasar outflows in the 500 -- 1050\angstrom\ rest-frame: II\\ The Most Energetic Quasar Outflow Measured to Date}
%\title{Bridging Optical Integral Field and UV Absorption Lines Analyses: \\ The AGN Outflows in Mrk 509 and IRAS F04250$-$5718}
%\title{The P\lowercase{a}$\alpha$ Luminosity Function of H \lowercase{{\sc ii}} Regions in Nearby Galaxies from HST/NICMOS$^{\ast}$}


\author{
Xinfeng Xu\altaffilmark{1,$\dagger$},
Nahum Arav\altaffilmark{1},
Timothy Miller\altaffilmark{1},
Gerard A. Kriss\altaffilmark{2},
Rachel Plesha\altaffilmark{2}
}

\affil{$^1$Department of Physics, Virginia Tech, Blacksburg, VA 24061, USA\\
$^2$Space Telescope Science Institute, 3700 San Martin Drive, Baltimore, MD 21218, USA
%\hspace{07mm}\
}

\altaffiltext{$\dagger$}{Email: xinfeng@vt.edu}

\begin{abstract}

We present a study of the BAL outflows seen in quasar SDSS J1042+1646 (z = 0.978) in the rest-frame 500 -- 1050\angstrom\ (EUV500) region. The results are based on the analysis of recent \textit{Hubble Space Telescope/Cosmic Origins Spectrograph} observations. Five outflow systems are identified, where in total they include $\sim$70 outflow troughs from ionic transitions. These include the first non-solar detections from transitions of \ov*, \nev*, \arvi, \cavi, \cavii, and \caviii. The appearance of very high-ionization species (e.g., \neviii, \naix, and \mgx) in all outflows necessitates at least two-ionization phases for the observed outflows. We develop an interactive Synthetic Spectral Simulation method to fit the multitude of observed troughs. Detections of density sensitive troughs (e.g., \siv*\ \ly 657.32\angstrom\ and the \ov*\ multiplet) allow us to determine the distance of the outflows ($R$) as well as their energetics. Two of the outflows are at $R$ $\simeq$ 800 pc and one is at $R$ $\simeq$ 15 pc. One of the outflows has the highest kinetic luminosity on record ($\dot{E_{k}}$ $ = 5\times 10^{46}$ erg s$^{-1}$), which is 20\% of its Eddington luminosity. Such a large ratio suggests that this outflow can provide the energy needed for active galactic nucleus feedback mechanisms.

\end{abstract}

%\keywords{galaxies: active -- galaxies: kinematics and dynamics -- quasars: absorption lines -- ISM: jets and outflows: general -- quasar: individual (SDSS J1042+1646)}
\keywords{galaxies: active -- galaxies: kinematics and dynamics -- quasars: jets and outflows -- quasars: absorption lines -- quasars: general -- quasars: individual (SDSS J1042+1646)}

\section{INTRODUCTION}
\label{Introduction}
Quasar absorption outflows are identified by blue-shifted troughs that appear in quasar spectra \cite[e.g.,][]{Hall02,Arav13,Grier15,Leighly18,Hamann19}. These outflows are believed to play a major role in various active galactic nucleus (AGN) feedback mechanisms (see elaboration in section 1 of Arav et al. 2019, submitted to ApJS, hereafter Paper I). Observations and analyses show that these outflows can have enough kinetic energy to be major contributors to AGN feedback \cite[e.g.,][]{Moe09,Dunn10a,Borguet13,Arav13,Chamberlain15a,Xu19}. To assess how effective outflows are at contributing to AGN feedback, theoretical models compare the kinetic luminosity of the outflow ($\dot{\text{E}_{k}}$) to the Eddington luminosity of the central black hole (L$_{\text{Edd}}$). These models predict that an Eddington ratio, i.e., $\Gamma_{\text{Edd}}$ $\equiv$ $\dot{\text{E}_{k}}$/L$_{\text{Edd}}$, of at least 0.5 -- 5\% is required for strong AGN feedback effects \cite[][respectively]{Hopkins10, Scannapieco04}. 

%: quenching of star formation and regulating the growth of supermassive black holes (SMBH) \cite[][]{Silk98,Ferrarese00,King03, Di05,Ostriker10, Hopkins10, Soker11, Hopkins16,Ciotti17,Angles17,Choi18}, explaining the observed shape of the quasar luminosity function \cite[][]{Hopkins05a,Hopkins07,Hopkins10b,FaucherGiguere12}, and chemically enriching the surrounding galactic medium \cite[][]{Moll07,McCarthy10,Baskin12,tay15}.

%Moreover, there are recent cosmological hydrodynamic simulations which explored the role of AGN feedback in the evolution of galaxies \cite[][]{Choi18}. They show that the AGN feedback can effectively quench the star formation and transform blue compact galaxies into red extended galaxies. 

Broad absorption line quasars (BALQSOs) are observed in $\simeq$ 20\% of the optically selected quasar population \cite[e.g.,][and references therein]{Hall02,Tolea02,Hewett03,Reichard03,Trump06,Ganguly08,Gibson09}. The majority of observed BALQSOs only show absorption troughs from high-ionization species, e.g., \civ, \nv, \siiv\ (and \hi), and are designated as HiBALs. There are more than 10,000 ground-based spectra of such outflows. However, almost all of these cover rest-frame wavelengths longer than 1050\angstrom\ \cite[][]{Arav13}. For these wavelengths, it is very rare to find diagnostic troughs that allow distance measurements of the outflow from the central source ($R$) \cite[e.g.,][]{Hamann98,Borguet12b, Borguet13, Capellupo14, Capellupo17, Arav18, Xu19}. Furthermore, the $\lambda_{\text{rest}} >$ 1050\angstrom\ cannot probe the very-high ionization phase (VHP, probed by troughs from \neviii, \mgx, \sixii, and higher-ionization species). The VHP has been shown to carry more than 90\% of the outflow's total hydrogen column density (\Nh) both in X-ray spectra of Seyfert outflows \cite[e.g.,][]{Behar17} as well as in quasar outflows where troughs from the above species were covered \cite[][]{Arav13,Finn14}. These two parameters ($R$ and \Nh) are the key to determining $\dot{\text{E}_{k}}$ (see equation (\ref{eq:2})) and, therefore, the possible contribution of the outflow to AGN feedback.

The 500 --1050\angstrom\ rest-frame region (hereafter EUV500) contains 10 times more measurable absorption features in a quasar outflow \cite[][]{Arav13}.   
Some of these have been observed and reported, including transitions from \niv, \oiv, \oiv*, \ov, \ovi, \neiv, \nev, \nevi, \neviii, \naix, \mgx, \alxi, \sixii, \sv, \arvii, and \arviii\ \cite[see table \ref{table:AtomicData} for details of the specific transitions;][]{Korista92, Telfer98, Telfer02, Arav01, Muzahid13, Finn14}. These include troughs from the very-high ionization phase, which can be used to determine \Nh\ \cite[][]{Arav13}, and troughs from more than 10 excited state transitions, which can be used to determine the electron number density (\ne) of an outflow and therefore $R$ \cite[e.g.,][]{Hamann01, Borguet12a, Xu18, Xu19}. Here we report the first non-solar detection of troughs from \ov*, \nev*, \arvi, \cavi, \cavii, and \caviii\ (see table \ref{table:AtomicData} for details). 
%In order to distinguish with XUV (1--912\angstrom) and Extreme-UV (EUV, 124--912\angstrom), we define this 500 -- 1050\angstrom\ rest-frame region as \XUV.  
%Most of the lines in \XUV\ have higher-ionization potential than \civ\ and \nv. 
%\niv\ \ly 765.147, \oiv\ \ly 608.40, \ly 787.71, \oiv*\ \ly 609.83, \ly 790.20, \ov\ \ly 629.73, \ovi\ \ly\ly 1031.91, 1037.61, \nev\ \ly 568.42, \nevi\ \ly 558.60, \neviii\ \ly\ly 770.41, 780.32,  \naix\ \ly\ly 681.72, 694.15, \mgx\ \ly\ly 609.79, 624.94,  \alxi\ \ly\ly 550.03, 568.12, \sixii\ \ly\ly 499.41, 520.66, \sv\ \ly 786.47, \arvii\ \ly 585.75 and \arviii\ \ly\ly 700.24, 713.80 
%\ov*\ (six transitions), \cav\ \ly\ly 637.91, 646.53, \cav*\ \ly 651.53, \cavi\ \ly\ly\ly 629.60, 633.84, 641.90 and  \cavii\ \ly 624.38 
%Unlike the Far-UV (FUV) region, the \XUV\ region has been largely unexplored, not because this region contains no useful absorptions from outflows, but because of the rarity of data for this range \cite[][]{Telfer02}. 

We carried out a \textit{Hubble Space Telescope (HST)} program GO 14777 (PI: Arav), which targeted a sample of 10 quasars (see selection criteria in Arav et al. 2019, submitted) and observed the \XUV\ with the \textit{Cosmic Origins Spectrograph (COS)}. Using HST is necessary since observing the \XUV\ with ground-based telescopes requires objects with redshift, z $\gtrsim$ 3. At these redshifts, the contamination from the \lya\ forest is severe, which does not allow for adequate measurements of \XUV\ troughs and, in most cases, not even their detections. Therefore, the EUV500 is only practicably accessible with the HST by observing bright quasars at 0.5 $\lesssim$ z $\lesssim$ 2 \cite[][]{Korista92, Arav99a, Telfer02, Arav01, Arav13, Finn14}.\\

%It has been shown for a long time that the multiplet of \ov*\ lines around 760\angstrom\ is useful density diagnostic transitions \cite[][]{Doyle83}. 

%Acceleration and deceleration of BALs remain an unsolved problem to date. The complications include: 1. the needs for long time baselines to see the accumulated acceleration signatures; 2. the difficulty to disentangle them from the velocity-dependent profile variability \cite[][]{Arav99a,Grier16}. There have been multiple studies of BAL variability by different groups \cite[][]{Gibson08,Gibson10,Capellupo12,Grier16}. However, they are mainly focused on studying it in FUV band by ground telescopes. Moreover, most frequently, the reported BAL acceleration appears in only a single transition and the $\Delta$v $\lesssim$ 1000 km s$^{-1}$. In this paper, we introduce an acceleration of outflow in one of the outflow \comps, for which the centroid shifted from -19500 to -21050 km s$^{-1}$. The acceleration is most obvious in \neviii\ doublets and is supported by other absorptions from \niv, and \mgx. This is the first time ever for BALs to have such a large acceleration appeared in multiple ion absorption troughs.

This paper is part of a series of publications describing the results
of  HST program GO-14777, which observed quasar outflows in the EUV500 using the COS.\\
Paper I \citep{ara20a} summarizes the results
for the individual objects and discusses their importance to various
aspects of quasar outflow research. \\
Paper II is this work.\\
Paper III \citep{mil20a} analyzes 4 outflows
detected in 2MASS J1051+1247, which show remarkable similarities, are
situated at $R\sim200$~pc and have a combined $\dot{E}_k=10^{46}$ erg
s$^{-1}$.  \\
Paper IV \citep{xu20b} presents the largest
velocity shift and acceleration measured to date in a BAL outflow.\\  
%Paper IV (Xu et al 2019c, submitted to ApJS) presents the largest
%velocity-shift and acceleration measured to date in a BAL outflow.  
Paper V \citep{mil20b} analyzes 2 outflows
detected in PKS 0352-0711, one outflow at $R=500$~pc and a second
outflow at $R=10$~pc that shows an ionization potential-dependent
velocity shift for troughs from different ions.\\ 
Paper VI \citep{xu20c} analyzes 2 outflows
detected in SDSS 0755+2306, including one at $R=1600$~pc with
$\dot{E}_k = 10^{46}$ -- 10$^{47}$ erg s$^{-1}$. \\
Paper VII \citep{mil20c} discusses the other
objects observed by program GO-14777, whose outflow characteristics
make the analysis more challenging.\\

%This is the first in a series of papers reporting the results from outflows with detected troughs shortwards of \oiv\ \ly 787.71 by HST program GO 14777. 
In this paper, we describe the analysis of outflows seen in quasar SDSS J1042+1646. The structure of the paper is as follows. In section 2, we discuss the details of the observations and data reductions. In section 3, we introduce our spectral analysis method. We present the analysis of each outflow \comp\ in section 4, where the photoionization solutions and \ne\ are determined. In section 5, we determine the distances and energetics for the outflows, and discuss these results in section 6. We summarize the paper in section 7. We adopt a cosmology with H$_{0}$ = 69.6 km s$^{-1}$ Mpc$^{-1}$, $\Omega_m$ = 0.286, and $\Omega_{\Lambda}$ = 0.714, and we use Ned Wright's Javascript Cosmology Calculator website \cite[][]{Wright06}.

\section{Observations and Data Reduction}
SDSS J1042+1646 (J2000: R.A. = 10:42:44.24, decl. = +16:46:56.14, z = 0.978 ) was observed with HST/COS \cite[][]{Green12} in November 2017 as part of our HST/COS program GO 14777 (PI: Arav). We used the G130M and G160M gratings, which have a resolving power of $\lambda/\delta\lambda\ \simeq\ $12,000 -- 17,000 and 14,000 -- 19,000, respectively \cite[][]{Fox18}. Our G130M observations cover the wavelength range of 1132 -- 1472 \angstrom\ and the G160M ones cover 1383 -- 1801\angstrom\ in the observed-frame. This object was observed previously in June 2011 using HST/COS G140L in program GO 12289 (PI: J. Howk). The COS G140L has a wider spectral coverage, 1100 -- 2000\angstrom\ and lower resolving power $\sim$\ 1500 -- 4000. There is a spectral gap in the G140L grating from around 1153\angstrom\ to 1185\angstrom. We show the details of both datasets in table \ref{table: epochs}.%The wavelength coverage varies depending on the choices of center wavelengths of the gratings

\begin{figure*}[htp]

\centering
	\includegraphics[trim={1cm 1.2cm 10.5cm 0.5cm},clip=true,width=1\linewidth,keepaspectratio]{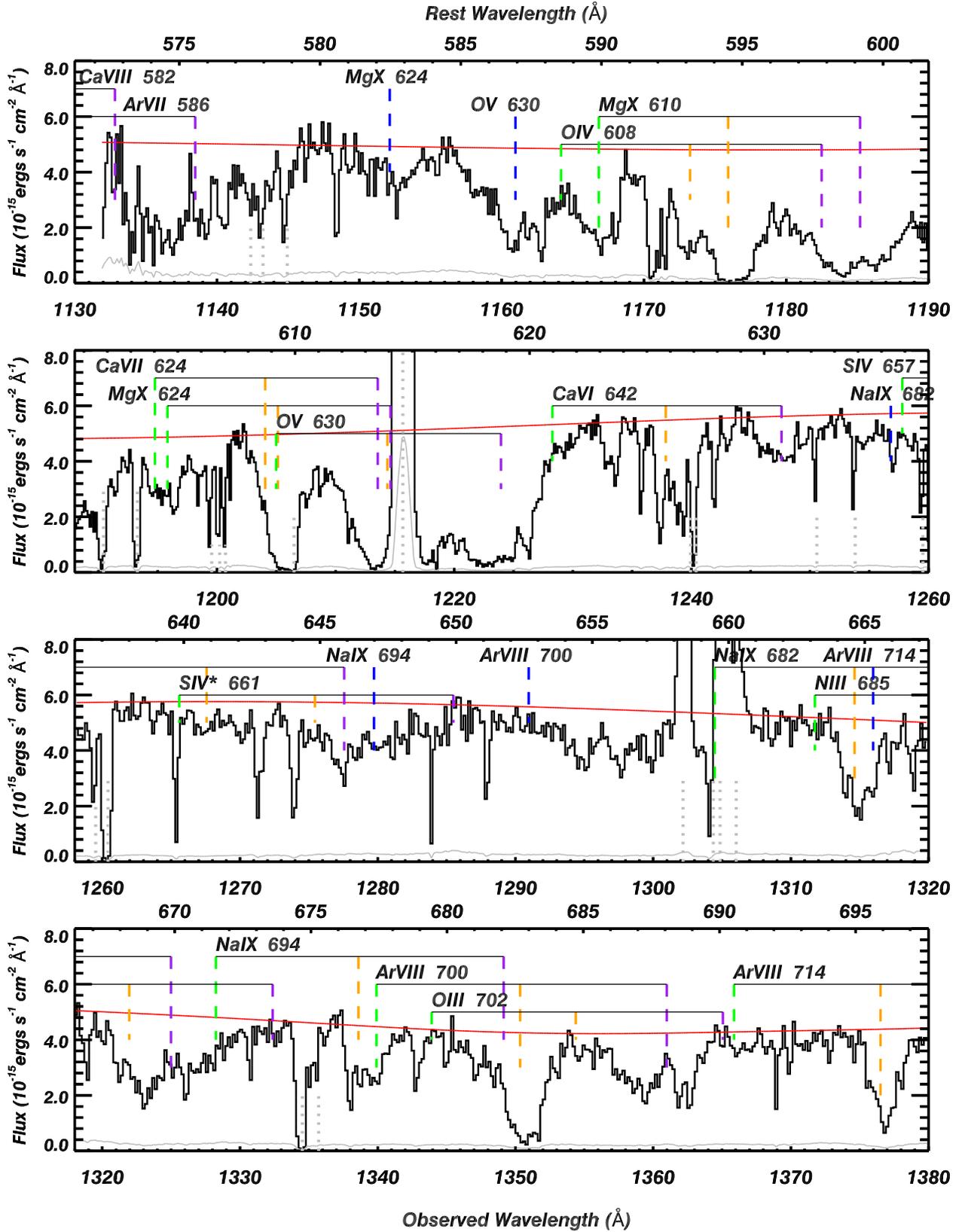}% trim: left lower right upper

\caption{HST/COS dereddened spectrum of SDSS J1042+1646 (z = 0.978). The black histogram shows the data from the 2017 epoch. The unabsorbed emission model and the flux error are shown as the red and gray solid lines, respectively. We label the important ionic absorption troughs in order of increasing velocity offset for all 5 outflow systems (see section \ref{text:AllComps}) with purple (S1a and S1b), orange (S2), green (S3), and blue (S4) vertical dashed lines. Strong Galactic ISM lines (e.g., \Siii\ at 1260.42\angstrom) and geocoronal lines (e.g., \hi\ at 1215.67\angstrom, \oi\ at 1302.17\angstrom, and \oi*\ at 1304.86\angstrom, 1306.03\angstrom) are marked with gray dotted lines under the spectrum. %The difference between the two epochs is mainly due to the flux change and calibration of the different gratings.
}

\label{fig:spec1}
\end{figure*}

We reduced the SDSS J1042+1646 data and estimated the errors following the same procedure described in \cite{Miller18}. For the 2017 observations, we combined the two observations for each grating. We corrected for Galactic extinction with E(B-V) = 0.022 \cite[][]{Schlafly12}. We show the full, dereddened spectrum in figure \ref{fig:spec1}.%, along with the spectrum from the 2011 epoch as a comparison.

\renewcommand{\thefigure}{\arabic{figure} (Continued)}
\addtocounter{figure}{-1}

\begin{figure*}[htp]

\centering
	\includegraphics[trim={1cm 1.2cm 10.5cm 0.cm},clip=true,width=1\linewidth,keepaspectratio]{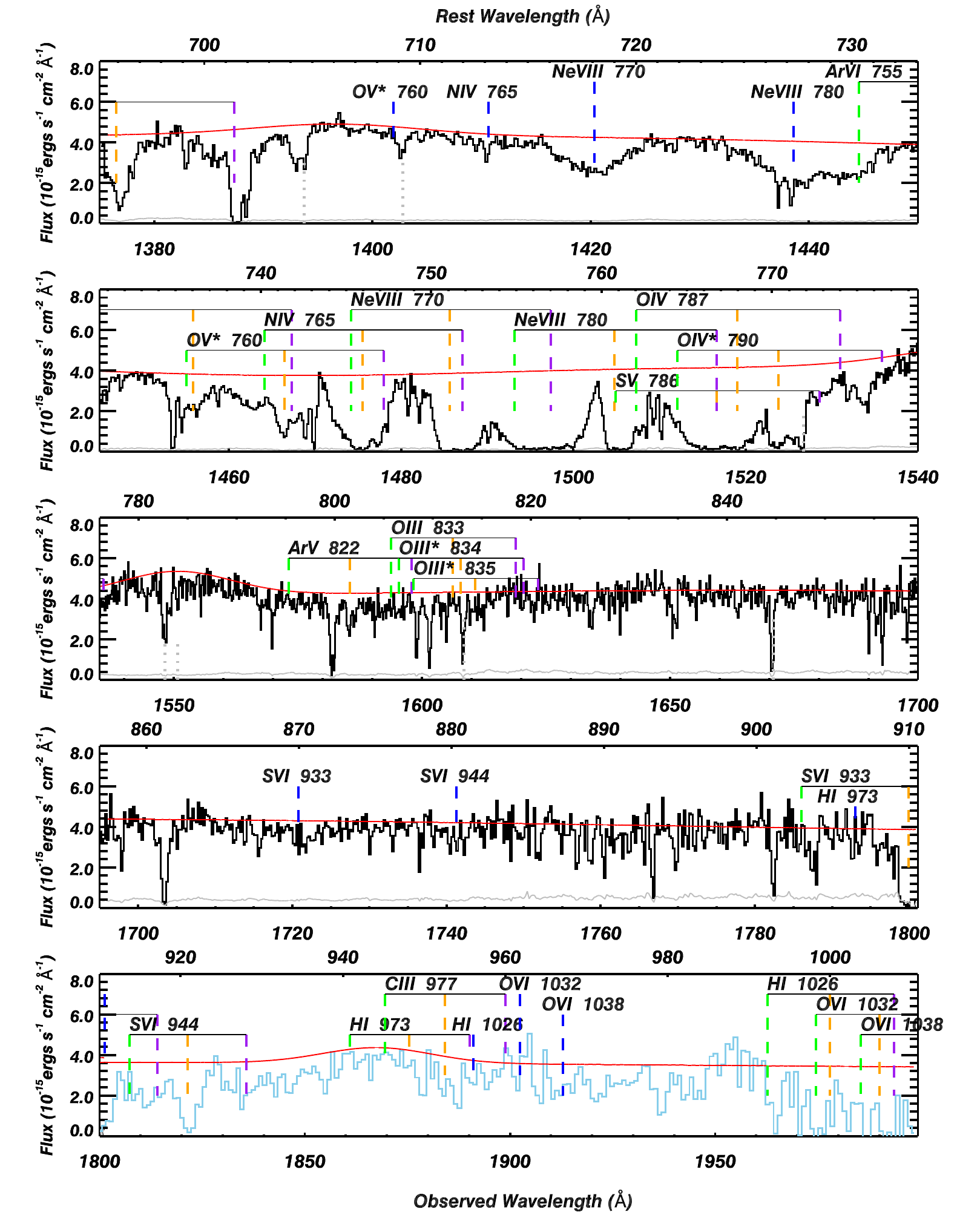}% trim: left lower right upper

\caption{In the last panel, we show the 2011 data in the blue histogram, which covers an extra range from 1800 -- 2000\angstrom. In other regions, the 2011 data are consistent with the 2017 data except for the highest velocity outflow (S4, marked by blue dashed lines), which we discuss elsewhere (Paper IV). 
%For outflow \comp\ 4, we observe an accelerations of outflows by 1550 km s$^{-1}$ from 2011 to 2017 epoch. We label the obvious velocity shift in \niv, \neviii, and \mgx. The blue and red vertical dashed lines are for 2011 and 2017 epochs, respectively. The red arrow under the \neviii\ \ly 770.41 represents the velocity shift size in wavelengths.
}
\label{fig:spec1:cont}
%(see section \ref{sec:comp4} for detailed discussion)
\end{figure*}

\renewcommand{\thefigure}{\arabic{figure}}

\begin{deluxetable}{c c c c}[htp]
%\tablewidth{0.4\textwidth}
%\tabletypesize{\small}
\setlength{\tabcolsep}{0.02in}
\tablecaption{HST/COS Observations for SDSS J1042+1646\label{table:epochs}}
\tablehead{
 \colhead{Epoch}	& \colhead{Date} & \colhead{Exp.(s)} 	& \colhead{Grating} 
}

\startdata
1 	& 2017 Nov 13, 18:20:01	&1720		&G130M			\\
2 	& 2017 Nov 13, 19:44:17	&1640		&G130M		\\
3 	& 2017 Nov 13, 21:08:53	&2320		&G160M			\\
4 	& 2017 Nov 13, 22:46:04	&2600		&G160M			\\
5 	& 2011 Jun 15, 19:17:32	&900		&G140L			\\

\vspace{-1.0mm}
\enddata

%\tablecomments{%The observation details of SDSS J1042+1646 by HST/COS.}
\label{table: epochs}
\end{deluxetable}

\begin{deluxetable}{ c c c}[htb!]
\tablewidth{0.5\textwidth}
\tabletypesize{\small}
\setlength{\tabcolsep}{0.02in}
\tablecaption{Outflows Detected in the SDSS J1042+1646 Data}
\tablehead{
 \colhead{Outflow \Comp}	& \colhead{Velocity$^{a}$ }	& \colhead{ \neviii\ Abs. Width$^{b}$} 			
\\
\\ [-2mm]
 \colhead{}  & \colhead{(km s$^{-1}$)} 	& \colhead{(km s$^{-1}$)}		
}

\startdata
\hline
S1			&	-5300 		&	2500			\\	
%System 1a		&	\textbf{-4950} 		&	1700			\\
%System 1b		&	\textbf{-5750} 		&	1700			\\
S2			&	-7500 		&	1500			\\
S3			&	-9940 		&	1350			\\
S4, 2011		&	-19500 		&	2000			\\
S4, 2017		&	-21050 		&	2000			\\

\vspace{-2.2mm}
\enddata

\tablecomments{
\\
$^{a}$ The velocity centroids come from the Gaussian profile fitting to unblended absorption troughs, e.g., \arviii\ \ly\ly 700.24, 713.80.\\
$^{b}$ \neviii\ \ly 770.41 absorption trough width is measured for continuous absorption below the normalized flux I = 0.9 \\
%(see discussion about the extended BAL definitions in section \ref{sec:BAL}).
%$^{b}$ Outflow \comp\ 1 has two components 1a and 1b, while for both of them the \neviii\ absorption trough widths are 1700 km s$^{-1}$.
}
\label{tb:OutflowSystems}
\end{deluxetable}

\renewcommand{\baselinestretch}{1.0}
\begin{deluxetable}{l c r c c c }[]
\tablewidth{0.45\textwidth}
%\tabletypesize{\small}
\setlength{\tabcolsep}{0.02in}
\tablecaption{Atomic Data for Observed Transitions in \XUV\ Range}
\tablehead{
 \colhead{Ion$^{(1)}$}	& \colhead{$\lambda$$^{(2)}$} 	& \colhead{\enspace\enspace E$_{low}$$^{(3)}$}	& \colhead{f$^{(4)}$} 		& \colhead{IP$^{(5)}$}		&\colhead{log(n$_{\text{e,crit}}$)$^{(6)}$}\\ 
 \colhead{}		& \colhead{(\AA)} 		& \colhead{\enspace\enspace(cm$^{-1}$)}		& \colhead{} 		& \colhead{(eV)}		&\colhead{log(cm$^{-3}$)}
 %\colhead{(1)}	& \colhead{(2)} 	& \colhead{\enspace(3)} 		& \colhead{(4)} 	& \colhead{(5)}		&\colhead{(6)}
}

\startdata
\hi			& 937.803	&0.00		&0.008		&13.60			&--\\
\hi			& 949.743	&0.00		&0.014		&13.60			&--\\
\color{red}\hi$^{(7)}$	& 972.537	&0.00		&0.029		&13.60			&--\\
\color{red}\hi		& 1025.72	&0.00		&0.079		&13.60			&--\\
\ciii			& 977.020	&0.00		&0.759		&47.89			&--\\
\niii			& 684.998	&0.00		&0.135		&47.45			&--\\
\niii			& 685.515	&0.00		&0.270		&47.45			&--\\
\niii*			& 685.817	&174.4		&0.320		&47.45			&3.3\\		
\niii*			& 686.336	&174.4		&0.069		&47.45			&3.3\\		
\niii 			& 763.334	&0.00		&0.084		&47.45			&--\\
\niii*			& 764.351	&174.4		&0.081		&47.45			&3.3\\
%\niii*			& 771.545	&57187.1	&0.146		&47.45			&--\\
%\niii*			& 771.901	&57246.8	&0.146		&47.45			&--\\
%\niii*			& 772.384	&57327.9	&0.146		&47.45			&--\\
%\niii*			& 772.889	&101023.9	&0.125		&47.45			&--\\
\niii			& 989.799	&0.00		&0.123		&47.45			&--\\
\niii*			& 991.511	&174.4		&0.122		&47.45			&3.3\\
%\niii*			& 991.511	&174.4		&0.012		&47.45			&--\\%combined to be one above
%\niii*			& 991.577	&174.4		&0.110		&47.45			&--\\
\color{red}\niv		& 765.147	&0.00		&0.611		&77.47			&--\\
\oiii*			& 599.590	&20273.27	&0.292		&77.47			&5.9\\%from PKS
\oiii			& 702.337	&0.00		&0.134		&54.93			&--\\
\oiii*			& 702.900	&113.2		&0.135		&54.93			&3.1\\
%\oiii*			& 702.838	&113.2		&0.045		&54.93			&--\\%combined to be one above
%\oiii*			& 702.896	&113.2		&0.035		&54.93			&--\\
%\oiii*			& 702.900	&113.2		&0.055		&54.93			&--\\
\oiii*			& 703.854	&306.2		&0.136		&54.93			&3.7\\
%\oiii*			& 703.850	&306.2		&0.034		&54.93			&--\\%combined to be one above
%\oiii*			& 703.854	&306.2		&0.102		&54.93			&--\\
\color{red}\oiii	& 832.929	&0.00		&0.106		&54.93			&--\\
\oiii*			& 833.749	&113.2		&0.106		&54.93			&3.1\\
%\oiii*			& 833.715	&113.2		&0.026		&54.93			&--\\%combined to be one above
%\oiii*			& 833.749	&113.2		&0.080		&54.93			&--\\
\oiii*			& 835.289	&306.2		&0.104		&54.93			&3.7\\%combined to be one above
%\oiii*			& 835.059	&306.2		&0.001		&54.93			&--\\
%\oiii*			& 835.092	&306.2		&0.015		&54.93			&--\\
%\oiii*			& 835.289	&306.2		&0.088		&54.93			&--\\
\color{red}\oiv		& 608.397	&0.00		&0.067		&77.41			&--\\
\color{red}\oiv*	& 609.829	&385.9		&0.067		&77.41			&4.1\\
\color{red}\oiv		& 787.710	&0.00		&0.111		&77.41			&--\\
\color{red}\oiv*	& 790.199	&385.9		&0.110		&77.41			&4.1\\
%\oiv*			& 790.112	&385.9		&0.011		&77.41			&--\\%combined to be one above
%\color{red}\oiv*	& 790.199	&385.9		&0.099		&77.41			&--\\
\color{red}\ov		& 629.732	&0.00		&0.512		&113.90			&--\\	
\ov* J=1		& 758.677	&82078.6	&0.080		&113.90			&--$^{(8)}$\\
\color{red}\ov* J=0	& 759.442	&81942.5	&0.191		&113.90			&--\\
\ov* J=1		& 760.227	&82078.6	&0.048		&113.90			&--\\
\color{red}\ov* J=2	& 760.446	&82385.3	&0.143		&113.90			&--\\
\ov* J=1		& 761.128	&82078.6	&0.064		&113.90			&--\\
\color{red}\ov* J=2	& 762.004	&82385.3	&0.048		&113.90			&--\\
\color{red}\ovi		& 1031.912	&0.00		&0.133		&138.12			&--\\
\color{red}\ovi		& 1037.613	&0.00		&0.066		&138.12			&--\\
\neiv 			& 541.128	&0.00		&0.055		&97.11			&--\\
\neiv 			& 542.074	&0.00		&0.110		&97.11			&--\\
\neiv 			& 543.884	&0.00		&0.170		&97.11			&--\\
\color{red}\nev 	& 568.424	&0.00		&0.110		&126.21			&--\\
\color{red}\nev* 	& 569.828	&411.2		&0.109		&126.21			&4.3\\
\color{red}\nev* 	& 572.335	&1109.5		&0.114		&126.21			&4.9\\
\nevi			& 558.603	&0.00		&0.092		&157.93			&--\\
\color{red}\neviii	& 770.409	&0.00		&0.102		&239.09			&--\\
\color{red}\neviii	& 780.324	&0.00		&0.050		&239.09			&--\\
\color{red}\naix	& 681.720	&0.00		&0.092		&299.90			&--\\
\color{red}\naix	& 694.150	&0.00		&0.045		&299.90			&--\\
\color{red}\mgx		& 609.793	&0.00		&0.084		&367.50			&--\\
\color{red}\mgx		& 624.941	&0.00		&0.041		&367.50			&--\\
\alxi			& 550.030	&0.00		&0.077		&442.00			&--\\	
\alxi			& 568.120	&0.00		&0.037		&442.00			&--\\
\sixii			& 499.406	&0.00		&0.072		&523.00			&--\\
\sixii			& 520.665	&0.00		&0.034		&523.00			&--\\
\color{red}\siv		& 657.319	&0.00		&1.130		&47.30			&--\\	
\color{red}\siv*	& 661.396	&951.4		&1.020		&47.30			&4.8\\
\siv*			& 661.455	&951.4		&0.118		&47.30			&4.8\\
\siv			& 744.904	&0.00		&0.249		&47.30			&--\\
\siv			& 748.393	&0.00		&0.459		&47.30			&--\\
\siv*			& 750.221	&951.4		&0.597		&47.30			&4.8\\
\siv*			& 753.760	&951.4		&0.131		&47.30			&4.8\\
\siv			& 809.656	&0.00		&0.118		&47.30			&--\\
\siv*			& 815.941	&951.4		&0.085		&47.30			&4.8\\
\color{red}\sv		& 786.468	&0.00		&1.360		&72.68			&--\\
\color{red}\svi		& 933.378	&0.00		&0.436		&88.05			&--\\
\color{red}\svi		& 944.523	&0.00		&0.215		&88.05			&--\\
\enddata
\label{table:AtomicData}
\end{deluxetable}

\addtocounter{table}{-1}

\begin{deluxetable}{l c r c c c }[]
\tablewidth{0.45\textwidth}
%\tabletypesize{\small}
\setlength{\tabcolsep}{0.02in}
\tablecaption{(Continued.)}
\tablehead{
 \colhead{Ion$^{(1)}$}	& \colhead{$\lambda$$^{(2)}$} 	& \colhead{\enspace\enspace E$_{low}$$^{(3)}$}	& \colhead{f$^{(4)}$} 		& \colhead{IP$^{(5)}$}		&\colhead{log(n$_{\text{e,crit}}$)$^{(6)}$}\\ 
 \colhead{}		& \colhead{(\AA)} 		& \colhead{\enspace\enspace(cm$^{-1}$)}		& \colhead{} 		& \colhead{(eV)}		&\colhead{log(cm$^{-3}$)}
}
\startdata
\ariv			& 850.605	&0.00		&0.043		&59.81			&--\\
\arv			& 705.353	&0.00		&0.061		&75.04			&--\\
\color{red}\arv		& 822.174	&0.00		&0.520		&75.04			&--\\
\arvi			& 544.730	&0.00		&0.218		&91.01			&--\\
\arvi			& 548.900	&0.00		&0.433		&91.01			&--\\
\color{red}\arvi	& 754.930	&0.00		&0.071		&91.01			&--\\
\arvi*			& 767.065	&2207.10	&0.488		&91.01			&6.1\\
%\arvi*			& 767.065	&2207.10	&0.440		&91.01			&--\\%combined to be one above
%\arvi*			& 767.729	&2207.10	&0.048		&91.01			&--\\
\color{red}\arvii  	& 585.748	&0.00		&1.210		&124.40			&--\\
\color{red}\arviii	& 700.240	&0.00		&0.375		&143.45			&--\\
\color{red}\arviii	& 713.801	&0.00		&0.180		&143.45			&--\\
\caiv			& 655.998	&0.00		&0.022		&67.15			&--\\%From PKS
\cav			& 637.917	&0.00		&0.014		&84.43			&--\\
\cav			& 646.534	&0.00		&0.040		&84.43			&--\\
\cav*			& 651.531	&3275.6		&0.053		&84.43			&6.3\\
\cavi			& 629.602	&0.00		&0.016		&108.78			&--\\	
\color{red}\cavi	& 633.844	&0.00		&0.032		&108.78			&--\\
\color{red}\cavi	& 641.904	&0.00		&0.049		&108.78			&--\\
\color{red}\cavii	& 624.385	&0.00		&0.058		&127.70			&--\\
\cavii*			& 639.150	&4071.4		&0.066		&127.70			&7.5\\%From PKS0350
\color{red}\caviii	& 582.845	&0.00		&0.078		&147.40			&--\\
\caviii*		& 597.935	&4308.3		&0.063		&147.40			&6.9\\%From PKS
\cax			& 557.765	&0.00		&0.326		&211.30			&--\\	
\color{red}\cax		& 574.010	&0.00		&0.160		&211.30			&--\\
\enddata

\tablecomments{\\
$^{(1)}$ Ions with one or more troughs detected in outflows observed in HST GO 14777. \\
$^{(2)}$ Rest wavelength of observed transitions. \\
$^{(3)}$ Lower-level energy of these transitions from the National Institute of Standards and Technology (NIST) database \cite[][]{NIST18}, except for \arvi*\ \ly 767.06, which is from \cite{Verner96}.\\
$^{(4)}$ Transition's oscillator strengths from the NIST database, except for \arv, \arvi, and \caiv\ -- \caviii, which are from \cite{Fischer06} (see details in Paper V).\\
$^{(5)}$ Ionization potential, which gives the energy required to ionize the element to the next stage of ionization  \cite[][]{Allen77}.\\
$^{(6)}$ The logarithm of the critical electron number density for the excited transitions. The n$_{\text{e,cirt}}$ values are from CHIANTI version 7.1.3 with assuming a temperature of 20,000 K \cite[][]{Landi13}.\\
$^{(7)}$ The ionic transitions that are observed in SDSS J1042+1646 are shown in \color{red}red\color{blk}.\\
$^{(8)}$ We discuss the \ov*\ multiplet in section \ref{sec:ne_comp2}.
%(9) \arvi*\ \ly 767.6 is not included in NIST.
}

\end{deluxetable}
\renewcommand{\baselinestretch}{\para}

\section{ Spectral Analysis}
\subsection{Unabsorbed Emission Model}
We model the unabsorbed emission of the two different epochs separately following the approach of \cite{Chamberlain15a, Miller18, Xu18}. The models include two components: 1) a continuum which is represented by a power law; 2) strong emission lines such as \ov\ and \neviii, which are fitted with one or more Gaussian profiles. 
The adopted emission model for the 2017 epoch is shown as the solid red line in figure \ref{fig:spec1}.

\subsection{Identifying Outflow Systems}
\label{sec:defineV}
The COS \XUV\ spectrum of SDSS J1042+1646 shows a variety of absorption features due to the interstellar medium (ISM) within our galaxy, intervening \lya\ forest systems, and quasar outflow absorption features at different velocities. We identify five kinematically distinct outflow systems (S1 -- S4, where S1 has two components: S1a and S1b, see table \ref{tb:OutflowSystems}). The full, dereddened spectrum is shown in figure \ref{fig:spec1}. For the 2017 epoch, we marked the strong absorption lines related to these five outflow systems with colored dashed lines (\comp\ 1a and 1b are merged for simplicity, see section \ref{text:comp1}). 
%Strong Galactic ISM lines are shown as gray dotted lines.

\subsection{Synthetic Spectral Simulation}
\label{sec:SSS}
% From equation (9) in \cite{Savage91}, the AOD ionic column density can be calculated from
%\begin{equation}\label{eq:sav91}
%\sub{N}{ion}=\frac{m_e c}{\pi e^2 f \lambda} \int \tau(v)dv
%\end{equation} 
%where $m_e$ is the mass of the electron, $c$ is the speed of light, $e$ is the electric charge, $\sub{N}{ion}$ is the ionic column density, $f$ is the oscillator strength, $\lambda$ is the wavelength, and $\tau(v)$ is the velocity dependent optical depth.
%The PC ionic column density can be determined from \cite[see, e.g.,][]{Arav05}
%\begin{equation}
%\sub{N}{ion}=\frac{m_e c}{\pi e^2 f \lambda} \int C(v)\tau(v)dv
%\end{equation}
%where $C(v)$ is the velocity dependent effective covering factor. 
% Also discussed in \cite{Arav05}, the PL ionic column density is
%\begin{equation}
%\sub{N}{ion}=\frac{m_e c}{\pi e^2 f \lambda} \int \frac{\tau(v)}{1+a(v)}dv
%\end{equation}
%where $a(v)$ is the power law index.} 
The main method to analyze AGN outflows involves the following steps: 1) identifying outflow systems at distinct velocities; 2) measuring ionic column densities (N$_{ion}$) of their absorption troughs using three standard methods, i.e., Apparent Optical Depth (AOD), Partial Covering (PC), and Power Law (PL); and 3) comparing these N$_{ion}$ with the predictions from photoionization models to determine the physical properties for each outflow \cite[e.g.,][]{Hamann01,Arav01, Arav12, Arav18, Borguet12a, Borguet13, Chamberlain15a, Xu19}. The AOD method assumes that at every velocity the outflow completely and homogeneously covers the emission source \cite[see, e.g.,][]{Savage91}. The PC method assumes a constant optical depth over a fraction of the emission source for every velocity \cite[see, e.g.,][]{Edmonds11}. The PL method assumes the outflow at each velocity completely covers the emission source but has a varying optical depth across the source in the form of a power law \cite[see, e.g.,][]{Arav05}.
	
For most of the previous works, the diagnostic absorption lines were within the \ly$_{rest}$\ $>$ 1050\angstrom. The above method works well in this range since there is a relatively small number of absorption troughs (mainly Ly$\alpha$, \civ, \nv, \siiv, and in rare cases \pv, and \siv). Blending of these absorption troughs between different outflow systems is minimal to moderate in most of the cases. 

%2) These transitions in the FUV are from ions with similar and relatively low ionization potential (IP), from IP(\siiv) = 45 eV to IP(\nv) = 98 eV. Thus, their N$_{ion}$ can usually be well fitted with a one-phase photoionization solution. 
%3) For the HiBALs, the only density diagnostic transitions in the range are the rarely seen \ciii* (1175\angstrom) and \siv* (1072\angstrom) \cite[][]{Borguet12b, Arav18, Xu19}. 
%Since most of the FUV objects published were observed from ground telescopes, they need a relatively high redshift (z $\simeq$ 2) to shift the FUV range into Atmospheric Window, thus the two density diagnostic transitions mentioned above suffer from the contaminations of the Ly$\alpha$ forest.\\

%Far-UV (FUV) band (1210 -- 2000\angstrom)
%Extreme-UV (EUV) band (100 -- 1210 \angstrom)
This analysis method is inadequate for the \XUV\ band since the \XUV\ band includes outflow troughs from $\sim$ 70 individual transitions. The atomic properties for these detected transitions are shown in table \ref{table:AtomicData} and the ionic transitions observed in SDSS J1042+1646 are marked in red. With more troughs and multiple outflow systems, trough blending is more severe than in the \ly$_{rest}$\ $>$ 1050\angstrom\ range, especially when the absorption troughs are wide [e.g., see the panel 2 in figure \ref{fig:spec1:cont}]. Therefore, it is difficult to measure N$_{ion}$ directly for most of the observed troughs in the \XUV. Similarly, it is impractical to measure upper limits for all possible transitions ($\sim$ 850 are within the wavelength range covered in these data alone).

%2) These transitions arise from species with IP ranging from $\sim$50 eV (e.g., \oiii, and \oiv) to as high as 500 eV [e.g. \neviii, \naix, \mgx, and \sixii\ \cite[][]{Gabel05, Arav13}]. As shown in \cite{Arav13}, these ion species %(e.g., \oiv\ and \mgx) cannot be explained by a one-phase photoionization solution. These characteristics of quasar outflows in the \XUV\ increase the analysis complexities and difficulties, requiring a new analysis method.  \\
%, including \oiv* (555\angstrom, 610\angstrom, 790\angstrom ), \ov* (multiplets around 760\angstrom), \nev* (570\angstrom, 572\angstrom), \siv* (661\angstrom, 750\angstrom, 816\angstrom), and \cav* (652\angstrom)
%2) There contains denser Ly$\alpha$ forest when moving to this shorter wavelength, which makes it hard to exclude the Ly$\alpha$ forest contaminations;

\begin{figure*}[htp]
\center
	\includegraphics[angle=0,trim={0.7cm 0.3cm 1.2cm 7.0cm},clip=true,width=0.85\linewidth,keepaspectratio]{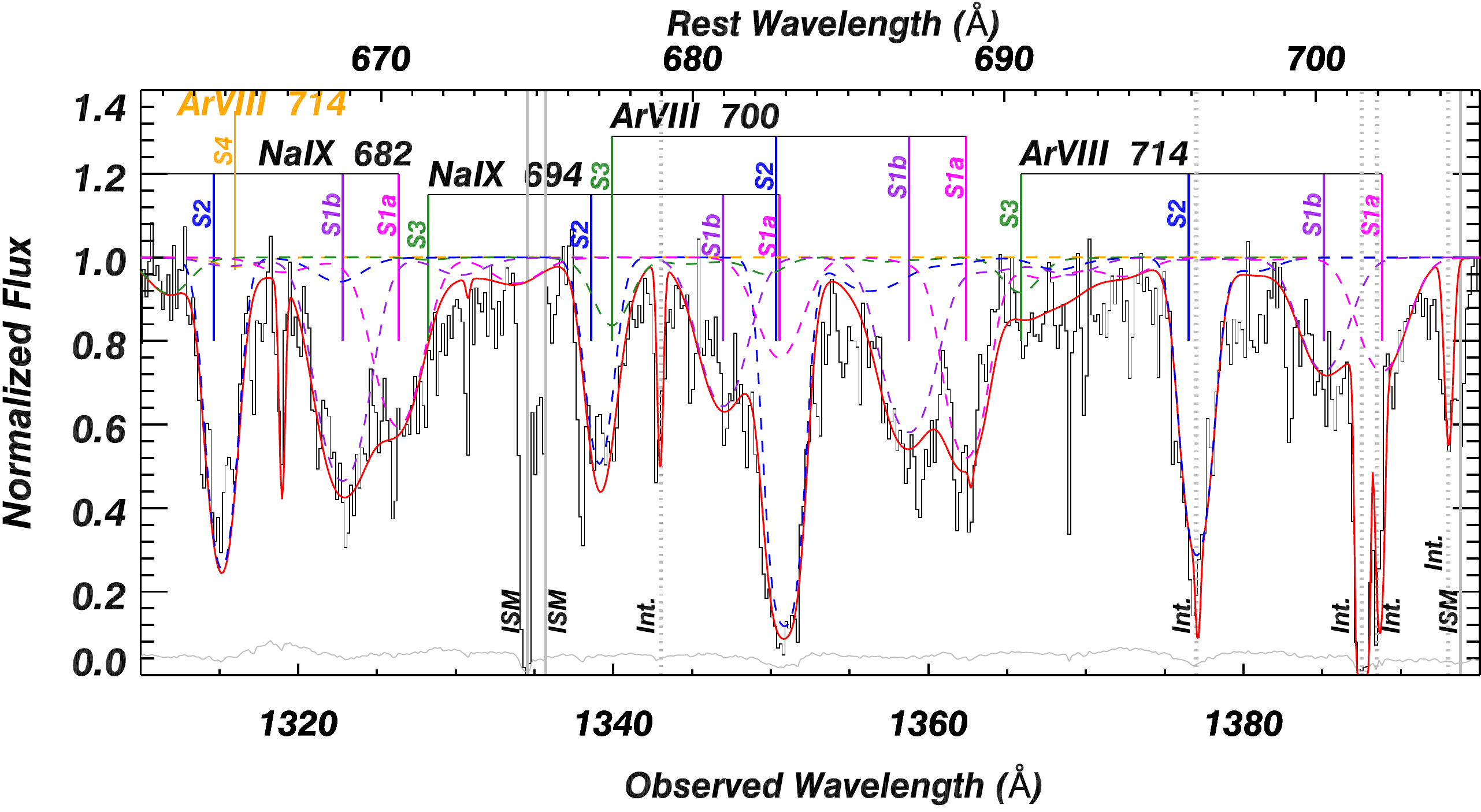}% trim: left lower right upper

\caption{Best-fitting \SSS\ models to the \arviii\ and \naix\ region. The black and gray histograms show the normalized flux and the corresponding error for the 2017 epoch, respectively. We label the five outflow \comps\ of the 2017 epoch and plot our models for them in magenta (S1a), purple (S1b), blue (S2), green (S3), and orange (S4) dashed curves (see section \ref{sec:defineV} for details). The solid red lines are the full \SSS\ model by adding these five outflow \comps. Strong Galactic ISM lines are shown in gray solid lines and marked with ``ISM" at the bottom of the plot. Similarly, we mark the known intervening systems by ``Int." and plot them as the gray dotted lines.}
% There are also strong intervening lyman systems, e.g. at 1387.5\angstrom, 1388.0\angstrom, and 1393.5\angstrom.

\label{fig:sample}
\end{figure*}

For these reasons, we developed a Synthetic Spectral Simulation (\SSS) method, which allows us to fit the multitude of troughs and determine the photoionization solutions more efficiently and accurately. While \SSS\ may appear to be superficially similar to the spectral synthesis code SimBAL \citep{Leighly18}, these two methods are philosophically quite different and to date have been applied to different rest-wavelength bands. The main analyzing steps of \SSS\ are:

(1) Instead of directly measuring \Nion\ of each absorption trough, we only measure the uncontaminated absorption troughs. For doublet transitions which are not heavily saturated, we obtain N$_{ion}$ measurements using the PC method as described in \cite{Arav08}. For saturated transitions, we use the AOD N$_{ion}$ and treat them as lower limits. For singlet transitions with a maximum optical depth, $\tau_{max}$, greater than 0.5, we treat the AOD N$_{ion}$ as lower limits. For absorption troughs with $\tau_{max}$ $<$ 0.05, we determine the AOD N$_{ion}$ as upper limits. For singlet transitions with 0.05 $<$ $\tau_{max}$ $<$ 0.5 and where the outflow shows other absorption troughs with $\tau>$ 2 from ionic transitions with similar ionization, we treat the singlets' AOD N$_{ion}$ as measurements since they are minimally affected by saturation effects ($<$ 10\%).  \\

%If there exist deeper absorption troughs with $\tau>$ 2,

(2) We then use these N$_{ion}$ as a basis to determine a preliminary photoionization solution (PI$_{1}$). Photoionized plasma in an AGN outflow is characterized by the total hydrogen column density, \Nh, and the ionization parameter, \Uh, where
\begin{equation}
\label{Eq:ionPoten}
\Uh=\frac{\Qh}{4\pi R^2 \nh c}
\end{equation}
%Therefore, we are able to solve for the distance, R, in equation \ref{Eq:ionPoten}.
$\Qh$ is the source emission rate of hydrogen ionizing photons, $R$ is the distance of the outflow from the central source, $\nh$\ is the hydrogen number density (for a highly ionized plasma, \nh\ $\simeq$ 0.8 \ne), and c is the speed of light.

%We simulate the observed spectrum based on the preliminary photoionization solution. 
We run the spectral synthesis code Cloudy [version c17.00, \cite{Ferland17}] to generate grids of photoionization simulations. We assume a spectral energy distribution (SED) and a metallicity \cite[e.g.,][and see elaboration in section \ref{subsection:comp1_PI}]{Arav13}. We vary log(\Uh) between -3.0 and 3.0, and log(\Nh) between 17.0 to 24.0 [hereafter, log(\Nion) is in units of log(cm$^{-2}$)] in steps of 0.05 dex. At each grid point, Cloudy predicts the N$_{ion}$ for all ions in its database. 
%For each combination of SED and metallicity, we only need to run this step once to get the grids (hereafter, $\textbf{G}$).

For each unblended absorption trough from a given ionic transition, we fit their optical depth profile in velocity space by a Gaussian profile as:

\begin{equation}
\label{Eq:GaussianFit1}
\tau(v)=\frac{A}{\sigma\sqrt{2\pi}}\times exp(\frac{(v-v_{c})^{2}}{2\sigma^{2}})
%I(v) = e^{-\tau(v)}
\end{equation}

where $\tau(v)$ is the optical depth profile of the absorption trough, $\mathit{v}$ is the velocity, A is the scaling factor, $\sigma$ is the velocity dispersion and v$_{c}$ is the velocity centroid of the Gaussian profile. $\frac{A}{\sigma\sqrt{2\pi}}$ is the maximum $\tau$ of the trough from the model. For each outflow system, we find the best v$_{c}$ and $\sigma$ by fitting the detected unblended absorption troughs. 

For each ion, $A$ in equation (\ref{Eq:GaussianFit1}) is scaled according to the predicted value of N$_{ion}$ based on PI$_{1}$ \cite[e.g.,][]{Savage91,Edmonds11}:

\begin{equation}
\label{Eq:GaussianFit2}
	\begin{split}
	N_{ion} &= \int {N(v)} dv  \\
		&=\int \frac{3.8 \times\ 10^{14}}{f\cdot\lambda}\cdot \frac{A}{\sigma\sqrt{2\pi}}\times exp(\frac{(v-v_{c})^{2}}{2\sigma^{2}})  dv
	\end{split}
\end{equation}

where $N(\mathit{v})$ is the velocity distribution of $N_{ion}$ (cm$^{-2}$ km$^{-1}$ s), $\mathit{f}$ is the oscillator strength of the corresponding ionic transition, and $\lambda$ is the rest wavelength of the transition. 

(3) We assume that all troughs in a given outflow system can be modeled using a Gaussian $\tau(v)$ function (see equation \ref{Eq:GaussianFit1}) with the same v$_{c}$ and $\sigma$ determined from the unblended absorption troughs of each outflow system. We vary $A$ to match the N$_{ion}$ predicted from PI$_{1}$, creating a full, AOD optical depth model for the entire observed spectrum. This model is then used to produce a synthetic spectrum which we overlay on top of the data. Since PI$_{1}$ has a minimal number of constraints (no constraints from blended troughs and maybe a few upper limits), there could be predicted troughs that conflict with the data. Therefore, we first look for predicted troughs that violate the $\tau _{max}$ upper limit threshold (discussed in step (1) above) within the continuum portion of the data so that we can measure additional upper limits. For example, PI$_{1}$ for S1a did not include the upper limit from \oiii. Inspection of the continuum region around the \oiii\ 833, \oiii* 834, and \oiii* 835 transitions marked in Figure~\ref{fig:spec1} showed predicted troughs from PI$_{1}$ that were not supported by the data ($\tau _{max}$ exceeded 0.05 for the strongest transitions at those wavelengths). Additional models are then created using \Uh\ and \Nh\ values that are on the boundary of the 1$\sigma$ contour for PI$_{1}$ and inspected to ensure that all upper limits are found.
	
(4) A new photoionization solution, PI$_{2}$, is determined for each outflow including any newly measured upper limits, and another synthetic spectrum is created. As an example, for S1a this resulted in a decrease in \Nh\ by 0.1 dex and an increase in \Uh\ by 0.03 dex for the high ionization phase (compared to PI$_{1}$) and no change in the very-high ionization phase (see section \ref{subsection:comp1_PI}). The blended troughs are then inspected for regions where none of the outflows have predicted troughs that can account for the observed absorption (none were found in these data). For such instances, the 1$\sigma$ contour of PI$_{2}$ is probed for all outflows like what was done is step (3), and if only one outflow trough is found that can match the absorption, a new lower limit (or in rare cases involving doublets, a measurement) can be made.

(5) A final solution (PI) is then created from all measurements and limits. As a final check, the unblended troughs yielding the measurements that produced PI$_{1}$ are visually inspected to ensure that there are not any contaminating troughs that would make the measurements unreliable.
%that visually matches the blended regions the best as our final solution, PI$_{1}$. 
As an example of the final results, we show the overall fit to the data (see figure \ref{fig:sample}), where we use the \SSS\ solutions derived in section \ref{text:comp1} -- \ref{sec:comp3}. The colored dashed lines are for different outflow systems, while the solid red lines are the overall model achieved by summing all outflow systems. \\
%\textbf{Add a better example?}\\
%, especially for the blended troughs regions, 
%Based on the N$_{ion}$, we then generate  to fit the absorption troughs.
%After that, based on a $\chi^{2}$-minimization of the difference between the model predicted troughs and the observed troughs, we adjust the photoionization solution until we get the best global fit. \\

\renewcommand{\baselinestretch}{1.0}
\begin{deluxetable}{c c l l}[htb!]
\tablewidth{0.5\textwidth}
\tabletypesize{\small}
\setlength{\tabcolsep}{0.02in}
\tablecaption{Column Densities for Outflow \Comps\ in SDSS J1042+1646}
\tablehead{
	 \colhead{Ion}		&\colhead{$\lambda$$^{(1)}$}	& \colhead{ N$_{ion,mea}$$^{(2)}$} 	 & \colhead{ $\frac{N_{ion,mea}}{N_{ion,model}}$$^{(3)}$  }   %& \colhead{ \frac{N$_{ion,mea}$}{N$_{ion,model}$}$^{(5)}$} 			
\\
\\ [-2mm]
	 \colhead{} 		& \colhead{(\AA)}		& \colhead{log(cm$^{-2}$)}	 	& \colhead{1}		%& \colhead{1}		
}

\startdata		
%\color{red} UL
%\color{blu} LL
%note: 	1. aod for all singlet, pc for uncontaminated doublets, need to check carefully
%	2. check the inconsistency between model and meas. Some of the saturated lines can be viewed as LL (such as NeVIII meas)
%	3. Redo the error size for PC solutions. They should not be always 0.1.
%	4. check all wavelengths, sometimes the wavelength I only use one of the doublets.
%	5. for OIV you need to only count the OIV not OIV+OIV*. 
% steps: first check consistency with table, including ion-wave, pc or aod results, model results.
\multicolumn{4}{l}{\textbf{Outflow \Comp\ 1a, v = [-5300,-4500]$^{(4)}$}}\\ 
\hline
			\hi		&972.54			&\color{blu}$>$15.32			&$>$0.09		\\
			%\hi		&1025.72		&\color{blu}$>$15.96			&16.59		\\	%1025 is close to the edge, so less reliable
			\oiii		&702.34			&\color{red}$<$14.80			&$<$0.77		\\	%UL fine
			\oiv	 	&787.71			&\color{blu}$>$15.19			&$>$0.05		\\	%UL fine, =OIV*+OIV
			\ov	 	&629.73			&\color{blu}$>$15.24			&$>$0.01		\\	%LL fine
			\neviii		&770.41, 780.32		&\color{blu}$>$16.40			&$>$0.06		\\	%LL fine
			\naix	 	&682.72, 694.15		&15.59$^{+0.15}_{-0.10}$		&1.00		\\	%682 clean, 694 left wing clean. PC fine (Need mention in text).
			\mgx	 	&609.79, 624.94		&\color{blu}$>$16.04			&$>$0.21		\\	%LL fine
			\siv	 	&657.32			&14.19$^{+0.10}_{-0.13}$		&1.11		\\	%Mea fine, =SIV*+SIV
			\sv	 	&786.47			&\color{blu}$>$14.04			&$>$0.42		\\	%LL fine
			\arv	 	&822.17			&\color{red}$<$15.21			&$<$5		\\	%UL fine
			\arvii	 	&585.75			&\color{blu}$>$14.48			&$>$0.53		\\	%LL fine, the trough shape is the same as NaIX and OV
			\arviii		&700.24, 713.80		&15.09$^{+0.13}_{-0.12}$		&0.91		\\	%PC fine
			\cavi	 	&633.84			&14.69$^{+0.12}_{-0.10}$		&1.00		\\	%UL fine
			\caviii		&582.85			&\color{blu}$>$15.43			&$>$0.50		\\	%LL fine	
\hline
\multicolumn{4}{l}{\textbf{Outflow \Comp\ 1b, v = [-6400,-5300]$^{(4)}$}}\\ 
\hline
			\hi		&1025.72		&\color{blu}$>$15.29			&$>$0.09		\\
			\niii		&685.52			&\color{red}$<$14.47			&$<$1.43	\\	%UL fine
			\oiii		&832.93			&\color{red}$<$14.81			&$<$0.77		\\	%UL fine
			\oiv	 	&787.71			&\color{blu}$>$15.59			&$>$0.12		\\	%LL fine
			\ov	 	&629.73			&\color{blu}$>$15.43			&$>$0.02		\\	%LL fine
			\neviii		&770.41, 780.32		&\color{blu}$>$16.38			&$>$0.04		\\	%LL fine
			\naix	 	&682.72, 694.15		&15.85$^{+0.15}_{-0.16}$		&1.11		\\	%682 clean, 694 left wing clean. PC fine (Need mention in text).
			\mgx	 	&609.79, 624.94		&\color{blu}$>$16.16			&$>$0.15	\\	%LL fine
			\siv	 	&657.32			&13.93$^{+0.15}_{-0.14}$		&0.91		\\	%mea fine, , =SIV*+SIV
			\sv	 	&786.47			&\color{blu}$>$14.76			&$>$2.00		\\	%LL fine
			\arv	 	&822.17			&\color{red}$<$15.18			&$<$5.00		\\	%UL fine
			\arvii	 	&585.75			&\color{blu}$>$14.73			&$>$0.83		\\	%LL fine, the trough shape is the same as NaIX and OV
			\arviii		&700.24, 713.80		&15.09$^{+0.12}_{-0.15}$		&0.83		\\	%PC fine
			\cavi	 	&633.84			&14.77$^{+0.12}_{-0.11}$		&1.11		\\	%UL fine
\hline
\multicolumn{4}{l}{\textbf{Outflow \Comp\ 2, v = [-8100,-6900]$^{(4)}$}}\\ 
\hline
			\hi		&1025.72		&\color{blu}$>$15.88			&$>$0.33		\\
			\niv		&765.15			&\color{blu}$>$15.34			&$>$0.91		\\
			\oiii	 	&702.34			&\color{red}$<$14.67			&$<$2.5		\\
			\oiv	 	&608.40			&\color{blu}$>$16.25			&$>$1.0		\\
			\neviii		&770.41, 780.32		&\color{blu}$>$16.43			&$>$0.06		\\
			\naix	 	&682.72, 694.15		&15.55$^{+0.16}_{-0.12}$		&0.95		\\
			\mgx	 	&609.79, 624.94		&\color{blu}$>$16.57			&$>$0.77	\\
			\siv	 	&657.32			&\color{red}$<$13.38			&$<$1.11		\\	
			\arv	 	&822.17			&\color{red}$<$15.60			&$<$10.0		\\
			\arviii		&700.24, 713.80		&15.30$^{+0.12}_{-0.10}$		&1.00		\\
			\cavi	 	&641.90			&14.90$^{+0.12}_{-0.15}$		&1.11		\\

\label{tb:IonSystems}
\end{deluxetable}

\addtocounter{table}{-1}

\begin{deluxetable}{c c l l }[htb!]
\tablewidth{0.5\textwidth}
\tabletypesize{\small}
\setlength{\tabcolsep}{0.02in}
\tablecaption{(Continued.)}
\tablehead{
	 \colhead{Ion}		& \colhead{ $\lambda$$^{(1)}$}		& \colhead{ N$_{ion,mea}$$^{(2)}$} 		 & \colhead{ $\frac{N_{ion,mea}}{N_{ion,model}}$$^{(3)}$  }  			
\\
\\ [-2mm]
	 \colhead{} 		& \colhead{(\AA)}		& \colhead{log(cm$^{-2}$)}	 		& \colhead{1}		
}

\startdata

\multicolumn{4}{l}{\textbf{Outflow \Comp\ 3, v = [-10500,-9500]$^{(4)}$}}\\ 
\hline	
			\hi		&1025.72		&\color{blu}$>$15.61			&$>$0.30		\\
			\niv		&765.15			&\color{blu}$>$14.84			&$>$0.26		\\
			\oiv	 	&608.40			&16.27$^{+0.12}_{-0.10}$		&1.00		\\
			\neviii		&770.41, 780.32		&\color{blu}$>$15.92			&$>$0.15		\\
			\naix	 	&682.72, 694.15		&\color{red}$<$15.68			&$<$10.0		\\
			\mgx	 	&609.79, 624.94		&15.51$^{+0.15}_{-0.14}$		&1.00		\\
			\arv	 	&822.17			&\color{red}$<$14.99			&$<$3.33		\\
			\arvi		&754.93			&\color{blu}$>$14.72			&$>$1.11		\\
			\arviii		&700.24, 713.80		&\color{red}$<$14.61			&$<$0.83		\\
			\cavi	 	&641.90			&\color{red}$<$14.95			&$<$1.43		\\
			\cavii	 	&624.39			&15.30$^{+0.16}_{-0.13}$		&1.11		\\		
\vspace{-2.2mm}

\enddata

\tablecomments{
\\
%$^{1}$ Measurable N$_{ion}$ and we don't include excited state transitions here since we examine them seperately for each outflow \comp\ in section \ref{text:AllComps}.\\
$^{(1)}$ Rest wavelengths of the measured transitions. For doublet or multiplet transitions, we show all the uncontaminated/measurable rest wavelengths. \\
$^{(2)}$ Measured N$_{ion}$ by PC or AOD method (see section 3.2). Lower limits are shown in {\color{blu}blue}, while upper limits are shown in {\color{red}red}.\\
$^{(3)}$ Ratio of the measured \Nion\ to the model predicted \Nion\ (see section 4) .\\
$^{(4)}$ N$_{ion}$ integration range in km s$^{-1}$.\\
%$^{6}$ \ov, \naix\ doublets, and \mgx\ \ly 624.94 fall into the gap of COS G140L grating for 2011 epoch (1152 -- 1185\angstrom).\\
%$^{7}$ \mgx\ \ly 609.79 is out of the observation range of COS G130M grating for 2017 epoch.
}

\label{tb:IonSystems2}
\end{deluxetable}
\renewcommand{\baselinestretch}{\para}

(6) To determine R, we use troughs from density sensitive transitions to obtain \ne\ \cite[e.g.,][]{Hamann01, Korista08, Borguet12a, Arav13, Xu19}. We vary the \ne\ of the outflow systems to fit the troughs of transitions from excited states (see section \ref{sec:ne_comp2} as an example). 

Overall, the \SSS\ is a three-dimensional model in the parameter space of \Uh, \Nh, and \ne. The detailed analysis for each outflow \comp\ is shown step by step in section \ref{text:AllComps}, and we discuss the strengths and caveats of this method in section \ref{sec:discussion}.\\
%We present our best-fitting \SSS\ model for the 1310 -- 1395\angstrom\ observed wavelength range in figure \ref{fig:sample}, which uses the photoionization solutions shown in section \ref{text:comp1} -- \ref{sec:comp3}.

\section {Analysis of Each Outflow System}
\label{text:AllComps}

In the five outflow systems, we observed strong absorption troughs from 1) high-ionization species, such as \niv, \oiii, \oiv, \ov, and \cavi, which have comparable ionization potential (IP) to the ionic transitions seen in HiBALs, e.g., \civ, \siv\ and \nv, from IP $\simeq$ 45 -- 100 eV. 2) very high-ionization species including \arviii, \neviii, \naix, and \mgx, with IP $\simeq$ 100 -- 500 eV. 3) transitions from excited states, e.g., \oiv*, \ov*, and \siv*, which are used to determine the \ne\ of the outflows. In our HST/COS spectra, we do not observe absorption troughs from any low-ionization species, which have IP smaller than \oiii\ (IP = 54.93 eV). Moreover, SDSS J1042+1646 was observed in 2006 by the Sloan Digital Sky Survey (SDSS)  in 1930 -- 4600 \angstrom\ rest-frame. We found no absorption troughs from the low-ionization ion transitions of \mgii\ \ly\ly 2796.35, 2803.53 (IP = 15.03 eV). Therefore, none of the outflows make SDSS J1042+1646 a low-ionization BALQSO \cite[][]{Voit93}.

The absorption troughs from outflow S1, S2, and S3 are consistent with no trough variability between the 2011 and 2017 epochs. However, S4 shows a velocity shift of --1550 km s$^{-1}$ between the two epochs. The remarkable behavior of S4 is described in Paper IV. In this paper, we take into account the blending of troughs from S4, which helps the analysis of the other three outflow \comps. 

Here, we report the detailed analyses of S1, S2, and S3. The N$_{ion}$ for the 2017 epoch are presented in table \ref{tb:IonSystems}. For each \comp, the measured column densities (N$_{ion,mea}$) summed over all observed energy states are reported in the third column. The ratios of measured to the modeled column densities (i.e., N$_{ion,mea}$/N$_{ion,model}$) from the \SSS\ method are reported in the fourth column. When we have N$_{ion,mea}$ as a lower limit, we expect N$_{ion,model}$ $>$ N$_{ion,mea}$, therefore, N$_{ion,mea}$/N$_{ion,model}$ $<$ 1 and vice versa.

%\comp\ 1, which is centered at v $\sim$ --5300 km$s^{-1}$ and can be divided further into two components as 1a and 1b at --4950 km $s^{-1}$ and --5750 km $s^{-1}$, respectively; \comp\ 2 at v = --7500 km $s^{-1}$; \comp\ 3 at v = --9940 km $s^{-1}$; and \comp\ 4 at v = --19500 km $s^{-1}$ for 2011 epoch and --21050 km $s^{-1}$ for 2017 epoch.

\subsection{Outflow \Comp\ 1 (v = --5300 km s$^{-1}$)}
\label{text:comp1}
\subsubsection{Kinematics and N$_{ion}$ Determinations}

\begin{figure}[htp]

\centering
	\includegraphics[angle=0,trim={0cm 0cm 0cm 7cm},clip=true,width=1\linewidth,keepaspectratio]{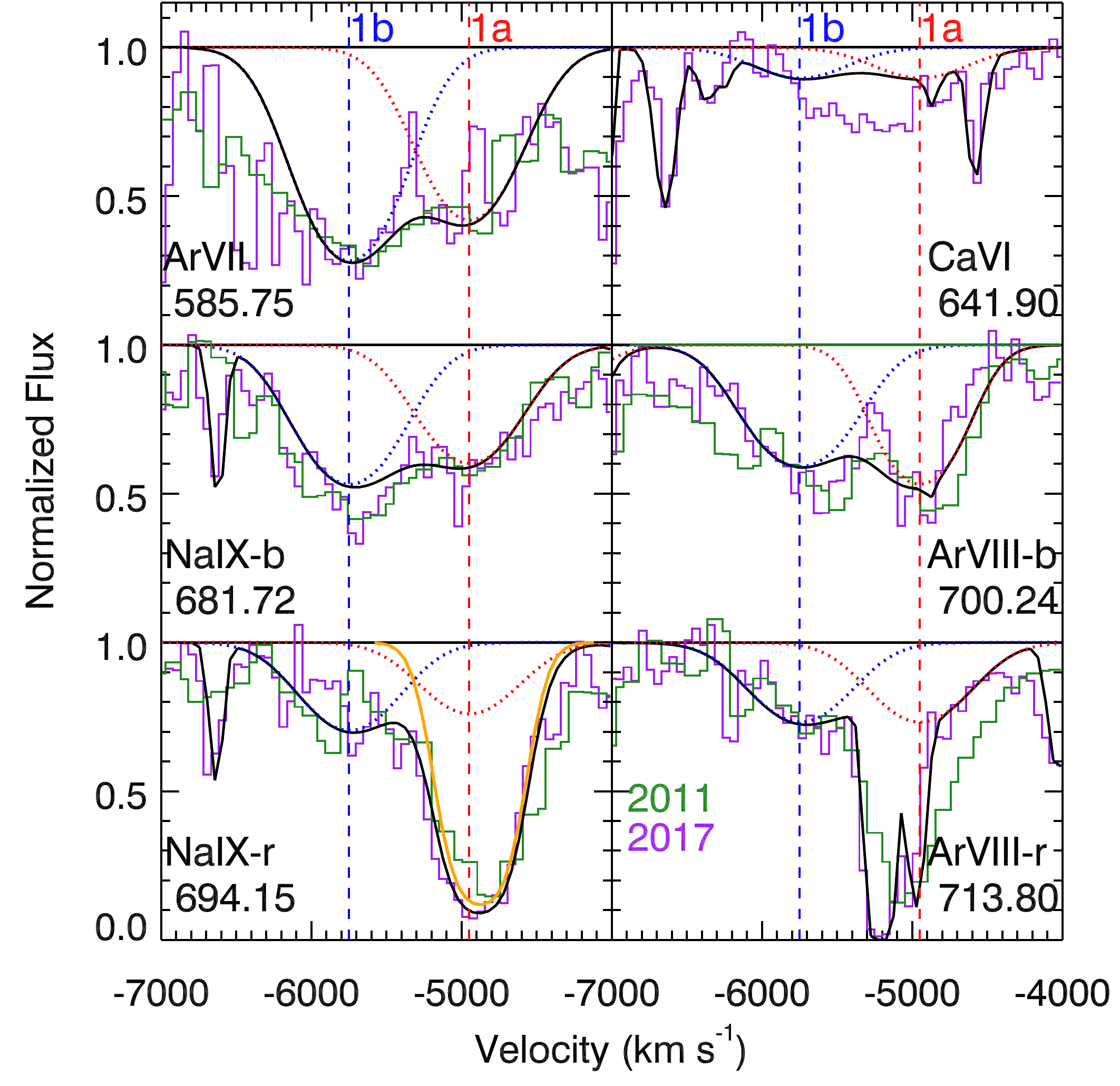}% trim: left lower right upper

\caption{Comparison between the data and \SSS\ models for the absorption troughs in components 1a and 1b of S1, where their velocity centers are marked by the red and blue dashed lines at -4950 km s$^{-1}$ and -5750 km s$^{-1}$, respectively. The normalized spectrum is shown in green for the 2011 epoch and purple for the 2017 one. The best-fitting \SSS\ models for components 1a and 1b are shown as red and blue dotted lines while the combined full model for each region is shown as the black solid lines. We mark the blue or red component of a doublet by adding a letter "b" or "r" after its ion label. In the \naix-r panel, the absorption from \arviii-b of outflow S2 is shown as the orange solid line. Strong intervening systems are modelled by narrow Gaussian profiles and are added to the full model, i.e., at v $\sim$ -6600 km s$^{-1}$ in the \naix-b and \naix-r region; v  $\sim$ -4950 and -5200 km s$^{-1}$ in the \arviii-r region; v $\sim$ -4600, -6350, and -6600 km s$^{-1}$ in the \cavi\ region.  }
\label{fig:vcut_comp1}
\end{figure}

S1 has strong absorption troughs from both high-ionization species, such as \oiii, \oiv, and \cavi, and very high-ionization species, including \naix, \mgx, and \neviii. In figure \ref{fig:vcut_comp1}, we show the kinematics of the absorption troughs for both the 2011 and 2017 epochs in green and purple, respectively. The troughs from --7000 km s$^{-1}$ to --4000 km s$^{-1}$ show double-minima features, which are most apparent in the \naix-b, \arvii\ and \arviii-b [hereafter, the ``-b" or ``-r" suffix stands for the shorter (bluer) or longer (redder) wavelength component of doublet transitions, respectively]. Since these two features appear at the same velocity in several troughs, we divide S1 into two components, 1a and 1b, which are centered at -4950 km s$^{-1}$ and -5750 km s$^{-1}$, respectively. For each individual ion, we report the measured N$_{ion}$ for the 2017 epoch in table \ref{tb:IonSystems}. \\

%\textbf{The model predicted absorption troughs from the \SSS\ method (see section \ref{subsection:comp1_PI}) are shown as the red and blue dotted lines for S1a and S1b, respectively. The overall model by summing up all components are shown as solid black lines.}\\

The N$_{ion}$ measurements used for the initial \Uh\ and \Nh\ solution are from the \naix\ and \arviii\ doublet transitions, and they are determined in the following ways: 1) For the \naix\ doublets in component 1b, we assume each trough is symmetric since the troughs are well fitted with Gaussian profiles (see figures \ref{fig:sample} \&\ \ref{fig:vcut_comp1}) and solve for N$_{ion}$ by using the PC equations on the unblended high-velocity wing. We then double the value to obtain the measurement. 2) Similarly, for the \arviii\ doublet in components 1a and 1b, the \arviii-r is contaminated by 2 intervening systems. Therefore, the unblended half of the troughs are used to solve the PC equations. Then, we double the values for the N$_{ion}$ measurements. 3) The \naix-r in component 1a is blended with the \arviii-b trough from S2. If no other information is available, we treat AOD N$_{ion}$ value derived from the \naix-b as a lower limit. However, the \arviii\ doublet of component 1a is not saturated and is deeper than the \naix-b\ absorption, while \arviii\ and \naix\ have similar IPs. Therefore, we can safely treat the AOD N$_{ion}$ value derived from the \naix-b as a measurement. \\

%For \neviii\ doublets and \ov, both the data and model are consistent with significant saturation. For 2011 epoch, the HST/COS G140L grating has a gap from around 1153\angstrom\ to 1185\angstrom, therefore the \cavi\ and \ov\ are not observed.

\indent Given the signal-to-noise (S/N) of the data and different spectral resolution of the observations, the absorption troughs are consistent with no variation between the two epochs for most of the ions. An exception is the low-velocity wing of the \arviii-r absorption trough. However, the entire \arviii-b troughs are consistent with no variations between the 2011 and 2017 epochs. Therefore, the deviation in \arviii-r is not intrinsic to the outflow system and may be due to calibration issues.

%describe siv Nion here.\\

%\indent The \oiv\ \ly 787.71 and \oiv*\ \ly 790.20 regions have a total of 4 absorption troughs from components 1a and 1b (see figure \ref{fig:ne_comp1}). The \oiv\  \ly 787.71 from component 1b (blue shaded region) is unblended, and the \oiv*\ \ly 790.20 of component 1a (yellow shaded region) is also uncontaminated and consistent with no absorption. Thus, we treat their AOD N$_{ion}$ as lower and upper limits, respectively. However, the \oiv\*  \ly 790.20 of component 1b and \oiv\  \ly 787.71 of component 1a are blended in the red shaded region, and we are unable to disentangle the N$_{ion}$ for each of them. Thus, we treat the AOD N$_{ion}$ summation of \oiv\ \ly 787.71 and \oiv*\ \ly 790.20 as lower limits for both components. 

%Moreover, \oiv\ and \oiv*\ near 788\angstrom\ seem to have moderate variations between 2011 and 2017 epochs. However, none of the other absorption troughs in outflow \comp\ 1 show similar variations. This change can come from the change of \ne\ between the two epochs. However, we report the N$_{ion}$ and \ne\ determinations based the 2017 epoch data since the grating G130M and G160M have much higher resolving power and S/N than the 2011 epoch's G140L grating. \\

\indent\cavi\ has three transitions at 629.60\angstrom, 633.84\angstrom, and 641.90\angstrom, where their $\mathit{f}$ ratios are almost 1:2:3, respectively. \cavi\ \ly 629.60 and \ly 633.84 are not measurable, since the former coincides with the strong \ov\ \ly 629.73 and the latter is blended with two intervening systems. However, we observed shallow absorption troughs from \cavi\ \ly 641.90, i.e, $\tau_{max}$ $<$ 0.2. Therefore, the AOD N$_{ion}$ derived will be close to the actual N$_{ion}$ value (see elaborations in section 2.9 \cite{Leighly11} and 5.4 of \cite{Arav13}). Thus, we treat the AOD N$_{ion}$ from the absorption trough of \cavi\ \ly 641.90 as a measurement. Similarly, for \siv\ \ly 657.32, we are able to measure the AOD N$_{ion}$ of it for both component 1a and 1b. We treat them as N$_{ion}$ measurements since they have $\tau_{max}$ $<$ 0.3. 

The strongest two observable transitions from \oiii\ are at 702.34\angstrom\ and 832.93\angstrom, and both are consistent with no absorption. We get an upper limit on N$_{ion}$ for \oiii\ from the AOD measurement of the stronger $\mathit{f}$ transition at 702.34\angstrom. Since oxygen is the most abundant metal, the non-detection of \oiii\ hints that none of the doubly ionized species would be strong enough to be detected in our spectrum. Indeed, the troughs from \niii\ are consistent with no absorption. \\
%This agrees with the high- and very-high ionization property of the detected outflows.\\
\indent\ The other ionic transitions are either saturated or undetected. Therefore, the AOD N$_{ion}$ for these two types are treated as lower and upper limits, respectively.

\begin{figure}[htp]

\centering
	\includegraphics[angle=0,trim={0cm 2.40cm 0.3cm 1.5cm},clip=true,width=1\linewidth,keepaspectratio]{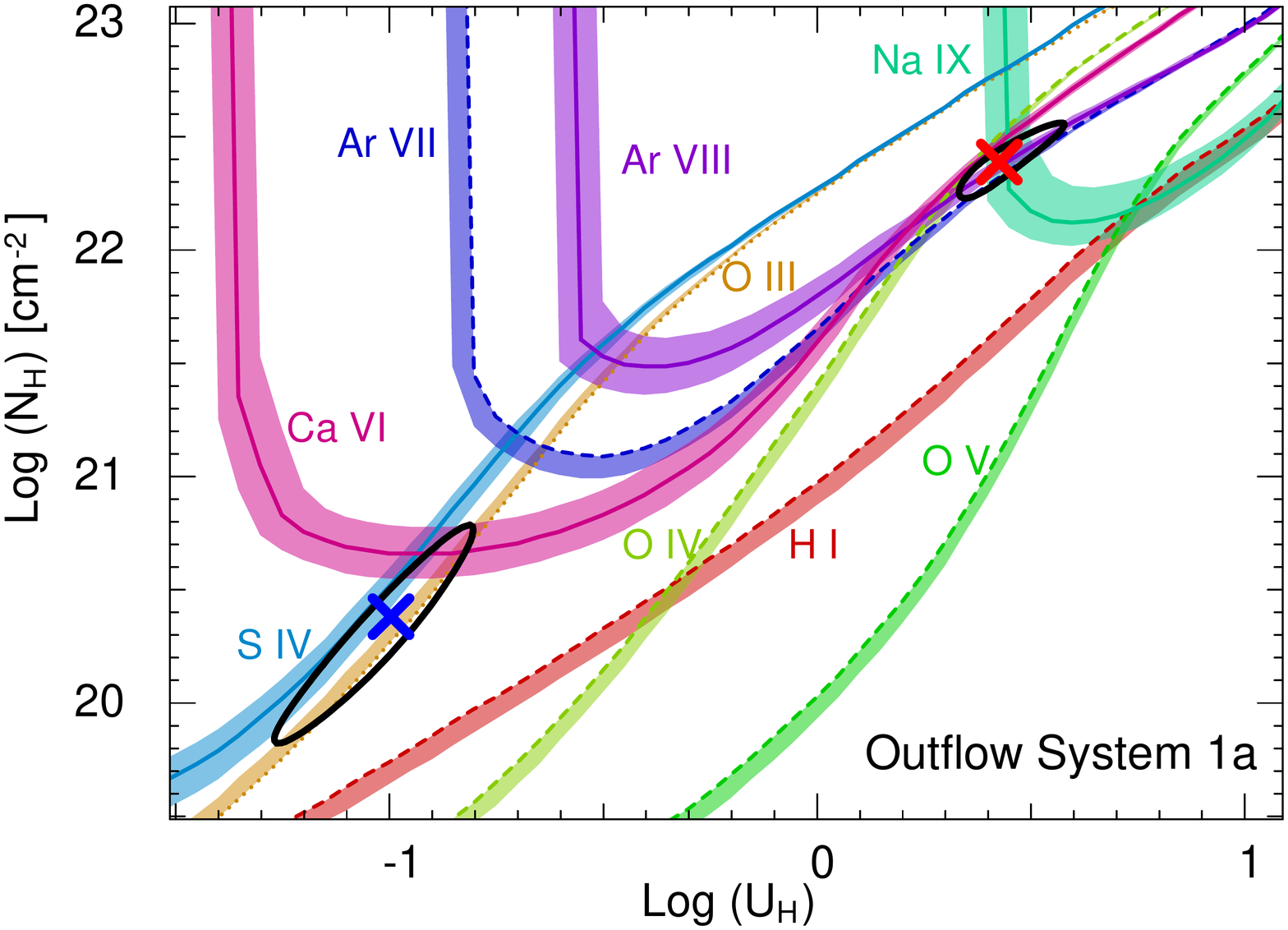} % trim: left lower right upper
	\includegraphics[angle=0,trim={0cm 0cm 0.3cm 1.70cm},clip=true,width=1\linewidth,keepaspectratio]{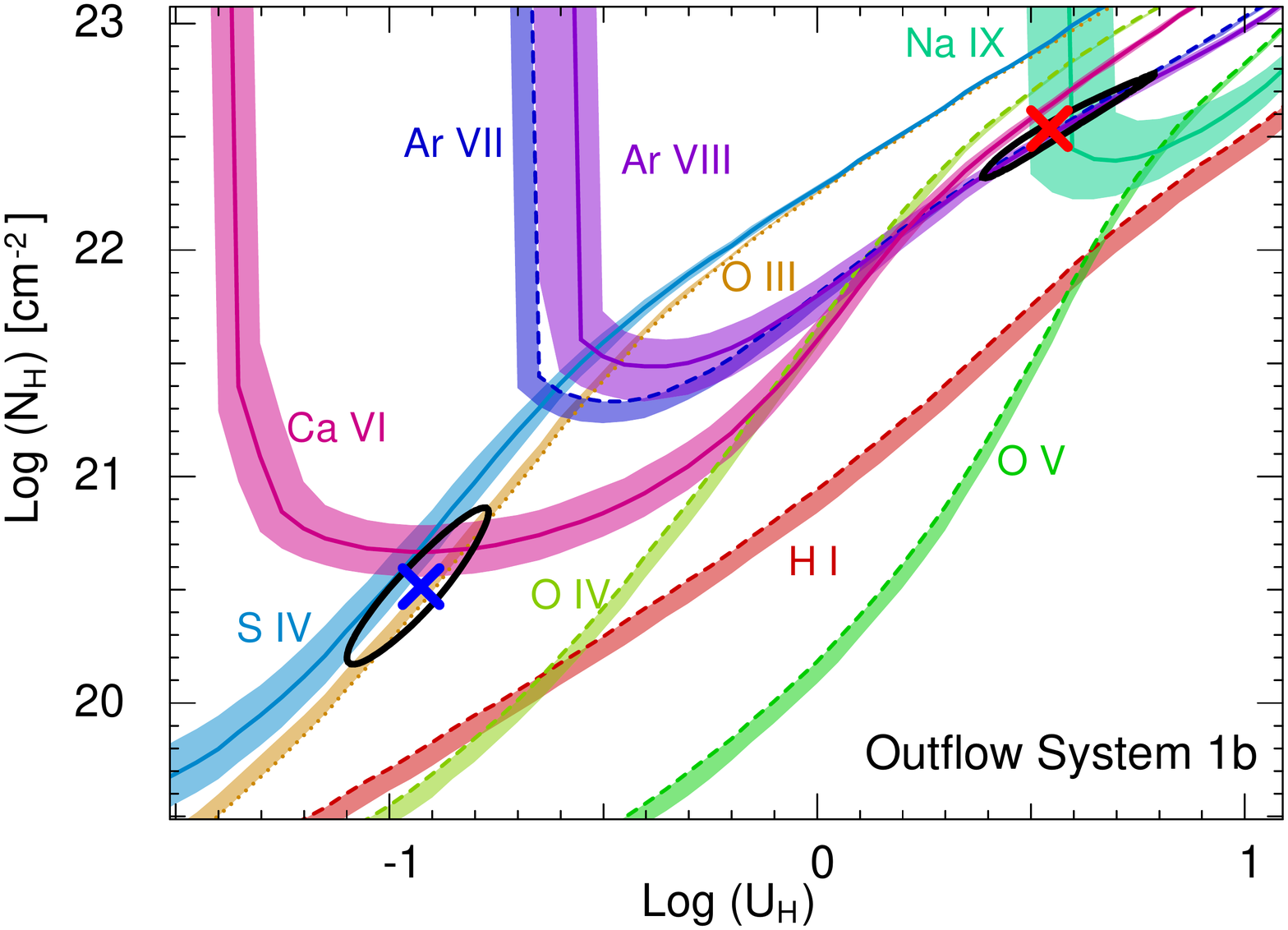} % trim: left lower right upper

\caption{The best-fitting photoionization solution for components 1a and 1b of S1 in epoch 2017. Each colored contour represents the region where the (\Nh, \Uh) model predicts consistent N$_{ion}$ with the observed ones. Solid lines represent N$_{ion}$ measurements, while dashed lines and dotted lines represent lower and upper limits, respectively. The very high- and high-ionization phase solutions are shown by the red and blue ``$\times$" along with their 1$\sigma$ error contours in the black ellipses, respectively. All other N$_{ion}$ in table \ref{tb:IonSystems} that are not shown here are lower or upper limits which are consistent with the solutions and are omitted for clarity's sake.}
\label{fig:comp1}
\end{figure}

\begin{figure}[htp]

\centering
	\includegraphics[angle=0,trim={0cm 0cm 4cm 11.5cm},clip=true,width=1\linewidth,keepaspectratio]{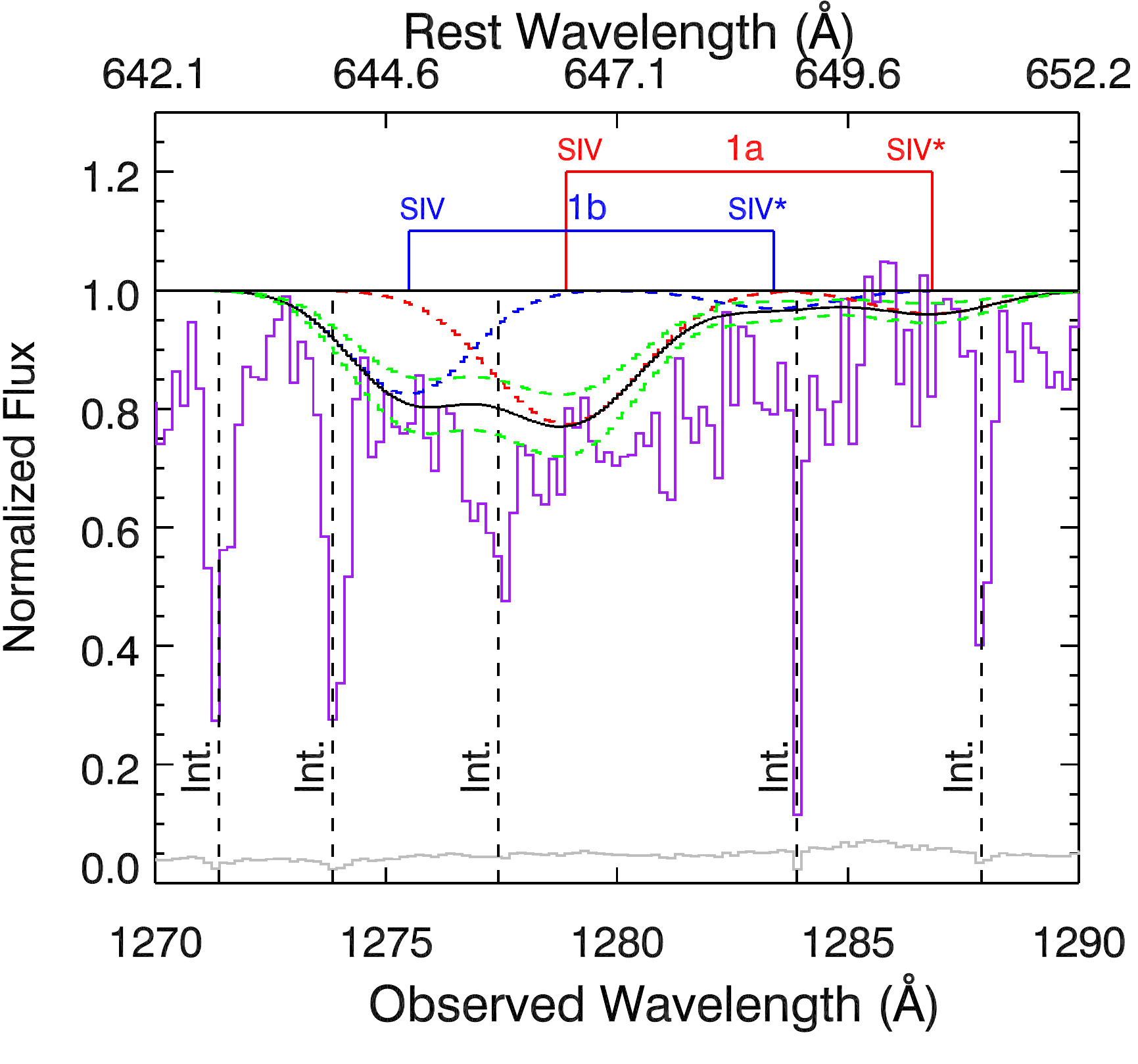}\\% trim: left lower right upper

\caption{\siv\ \ly 657.32 and \siv*\ \ly 661.40 regions for components 1a and 1b from S1. The purple and gray solid histograms are the normalized flux and errors for the 2017 epoch, respectively. The red and blue lines are for component 1a and component 1b, respectively. Based on our best-fitting photoionization solution, we show the predicted absorption troughs of \siv\ \ly 657.32 and \siv*\ \ly 661.40 in red and blue dashed lines. The full absorption model is shown as the solid black line, while strong intervening systems are marked with black dashed lines. We increase and decrease the optical depth such that the change in flux at the centers of the Gaussian profiles equal the noise level at that wavelength and adopt them as the lower and upper bounds in fitting the troughs, respectively. These bounds are shown in green dashed lines.
%\textbf{Bottom:} The \siv*/\siv\ column density ratio versus \ne\ from Chianti 7.1.3 atomic database \cite[][]{Landi13}. In order to show the dependency of \ne\ to temperature, we include curves for three different temperatures.  The red cross is the solution and the corresponding errors for component 1a ,where T = 15,000 K and is derived from the low-ionization phase of component 1a (see discussion in section \ref{text:ne_determine_comp1}).
}
\label{fig:ne_comp1}
\end{figure}

\subsubsection{ \SSS\ Models and Photoionization Solutions}
\label{subsection:comp1_PI}
After we determined N$_{ion}$ for unblended troughs, we follow the \SSS\ method in section \ref{sec:SSS} to get the best-fitting photoionization model for all observed troughs. To account for systematics in the unabsorbed emission model, the adopted N$_{ion}$ lower limits assume lower errors of 20\%, and N$_{ion}$ upper limits assume upper errors of 20\%. We assume solar metallicity \cite[][]{GASS10} and the HE0238 SED \cite[][]{Arav13}. This SED is based on observations of quasar HE0238--1904 in a similar rest-frame wavelength range to our object \cite[][]{Arav13}. For the observed data (570--1000 \AA\ rest-frame), the ratio of the HE0238 SED with the SDSS J1042+1646 continuum is constant to within $\pm$10\%. The best-fitting model is shown as the black solid lines in figure \ref{fig:vcut_comp1}. Absorption troughs from components 1a and 1b are plotted separately as blue and red dotted lines. The ratios of measured N$_{ion}$ to the model predicted N$_{ion}$ are presented in the 4th column in table \ref{tb:IonSystems}.

\cite{Arav13} demonstrated that detections of both the very high-ionization troughs (e.g., \neviii\ and \mgx) and high-ionization troughs (e.g., \oiv)  in the same outflow state two ionization phases for the absorber. We find the same situation in all of the outflow systems discussed here. Figure \ref{fig:comp1} shows the models' corresponding photoionization solution. For component 1a (see top panel of figure \ref{fig:comp1}), the very high-ionization phase (VHP) is marked as the red cross and is mainly constrained by the N$_{ion}$ measurements of \naix\, and \arviii. However, this VHP underestimates the observed N(\siv) by over a factor of 1000. Moreover, if we shift this VHP vertically up to match the N(\siv) curve at log(\Nh) = 22.78, the solution will overpredict the N(\arviii) by a factor of 13 and N(\cavi) by a factor of 22. Therefore, a pure VHP solution is unable to produce all of the observed \Nion. The high-ionization phase (HP) is marked by the blue cross and produces the observed N(\siv) and is constrained by the \Nion\ upper limit of \oiii. However, this HP produces negligible amounts of the observed N(\arviii) and N(\naix). Therefore, we invoke a two-phase ionization solution. The very high-ionization phase (VHP) has log(\Nh) = 22.37$^{+0.19}_{-0.12}$, and log(\Uh) = 0.42$^{+0.16}_{-0.09}$, and the high-ionization phase (HP) has log(\Nh) = 20.39$^{+0.40}_{-0.58}$ and log(\Uh) = --0.98$^{+0.17}_{-0.30}$. The N$_{ion}$ upper limits from \niii, \sv, and \arv, and the lower limits from \neviii, \mgx, and \caviii\ are consistent with this two phase solution and omitted for clarity's sake. 

Similarly, component 1b (see the bottom panel of figure \ref{fig:comp1}), has a VHP at log(\Nh) = 22.51$^{+0.27}_{-0.21}$ and log(\Uh) = 0.54$^{+0.24}_{-0.16}$, and a HP at log(\Nh) = 20.52$^{+0.36}_{-0.34}$ and log(\Uh) = --0.93$^{+0.17}_{-0.18}$. We discuss and compare the two-ionization phase solutions to former studies in section \ref{sec:2Phases}.

%The error sizes of the low-ionization phases are larger. This is not surprising as there is no N$_{ion}$ measurements for the low-ionization phase.  \\

\subsubsection{Determinations of \ne\ From \siv\ and \siv*}
\label{text:ne_determine_comp1}
The most robust way for determining the distance, $R$, from the ionizing source of an absorption outflow is to use the troughs from ionic excited states. The ratio between the N$_{ion}$ from excited and ground states, e.g., N(\siv*)/N(\siv), yields \ne\ for an outflow, and combined with the value of \Uh, $R$ can be determined \cite[See equation (\ref{Eq:ionPoten}). e.g.,][]{Borguet13, Chamberlain15b, Arav18, Xu18, Xu19}. 
%In S1, We analyze the density-diagnostic ionic transitions for component 1a and 1b separately here.\\

%By using ground telescopes, the resonance state of \siv\ at 1062.66\angstrom\ and the excited state at 1072.96\angstrom\ are one of the useful density diagnostic lines \cite[][]{Borguet12b, Arav18, Xu19}. But the low oscillator strength of this set of \siv\ (f = 0.0487) and \siv*\ (f = 0.0434) make them relatively rare to be detected. 

For component 1a, we observed the absorption troughs from the strongest \siv\ transition at  657.32\angstrom\ and the corresponding excited state at 661.40\angstrom\ both with oscillator strengths, $\mathit{f}$ $\sim$ 1.18. For the 2017 data,  we show our best-fitting \SSS\ model for the \siv\ region in figure \ref{fig:ne_comp1}. We clearly have a deeper feature in the \siv\ \ly 657.32 region than in the \siv*\ \ly 661.40 region for component 1a. This yields log[N(\siv)] = 14.19$^{+0.10}_{-0.13}$ and log[N(\siv*)] = 13.35$^{+0.11}_{-0.12}$. Following the same method described in section 2 of \cite{Xu19}, we derive log(\ne) = 3.68$^{+0.18}_{-0.25}$ for component 1a [hereafter, log(\ne) is in units of log(cm$^{-3}$)]. 
%Since the \siv*\ energy level is mainly populated by collisional excitation, \ne\ is related to the N(\siv*)/N(\siv).
%For 2011 data, these two ionic transitions are on the edge of the G140L grating's gap and are not measurable.
%This leads to the fact that the \ne\ of component 1a is less than the critical density of \siv*/\siv, which gives log(\ne) $<$ 4.7 [hereafter, log(\ne) is in units of log(cm$^{-3}$)] for T = 15000 K. This temperature is derived from the low-ionization phase solution of component 1a discussed in section \ref{subsection:comp1_PI}.

%We observed \oiv\ \ly 787.71 and \oiv*\ \ly 790.20 absorption troughs from both component 1a and 1b. Since the velocity separation between these two components is close to the separation of \oiv\ and \oiv*, component 1a and 1b blends at around 1533\angstrom\ observed wavelength. However, for component 1b, we have unblended absorption troughs from \oiv\ \ly 787.71, and the upper limit trough from \oiv*\ \ly 790.20 (see the bottom panel of figure \ref{fig:ne_comp1}). In this case, the \ne\ of component 1b is less than the critical density of \oiv*/\oiv\ and this yields the log(\ne) $<$ 4.1 for T = 15000 K.

Similarly, for component 1b, the absorption trough from \siv\ \ly 657.32 shows a deeper feature than \siv*\ \ly 661.40, while the latter cannot be much deeper than the current modeling due to the flux level on the two shoulders (see figure \ref{fig:ne_comp1}). We get log[N(\siv)] = 13.93$^{+0.14}_{-0.13}$ and log[N(\siv*)] = 13.21$^{+0.10}_{-0.13}$. This leads to log(\ne) = 3.78$^{+0.23}_{-0.26}$, which is consistent with the \ne\ result for component 1a within the errors. This strengthens the possibility that both outflows originate from the same absorber. After we determine \ne\ for the outflows, $R$ can be derived (see section \ref{sec:energy}).

\subsection{Outflow \Comp\ 2 (v = --7500 km s$^{-1}$)}
S2 also shows absorption troughs from both high- and very high-ionization species. Moreover, S2 has deep absorption troughs near its expected \ov*\ multiplet region. We begin by determining the N$_{ion}$ and constructing the photoionization solutions. Following that, we model the \ov*\ regions and derive the constraints for \ne\ and R.

\subsubsection{Kinematics and N$_{ion}$ Determinations}
\label{sec:S2:kine}
We show the strong absorption troughs for S2 in figure \ref{fig:vcut_comp2}, where the 2011 and 2017 epochs are shown in green and purple colors, respectively. The data is consistent with no variability between the 2011 and 2017 epochs. We report the measured N$_{ion}$ in table \ref{tb:IonSystems} and discuss them in detail here.

Similar to S1, we adopt the PC method and get N$_{ion}$ measurements from the unblended and well-separated \naix\ and \arviii\ doublets. The \neviii\ and \mgx\ doublet troughs are black at their deepest points, so their AOD N$_{ion}$ from the red component are taken as lower limits. For the \cavi\ triplet, the absorption trough from \cavi\ at 629.60\angstrom\ and 633.84\angstrom\ are blended with the Galactic \lya\ region, while the \cavi\ \ly 641.90 shows similar but shallower trough features as the \naix\ doublet. After modeling out the contamination of intervening systems (i.e., at v = --7050, --7320, --7720 km s$^{-1}$) by narrow Gaussian profiles, the AOD N$_{ion}$ value from the \cavi\ \ly 641.90 is treated as a measurement. The absorption troughs from other ion transitions are either lower or upper limits and are reported in table \ref{tb:IonSystems}. The lack of detection from \oiii\ and lower IP ions demonstrates that this is not a low-ionization BALQSO.\\

\subsubsection{\SSS\ Models and Photoionization Solutions}
\label{sec:OS2:PI}
\begin{figure}[htp]

\centering
	\includegraphics[angle=0,trim={0cm 0cm 0cm 0cm},clip=true,width=1\linewidth,keepaspectratio]{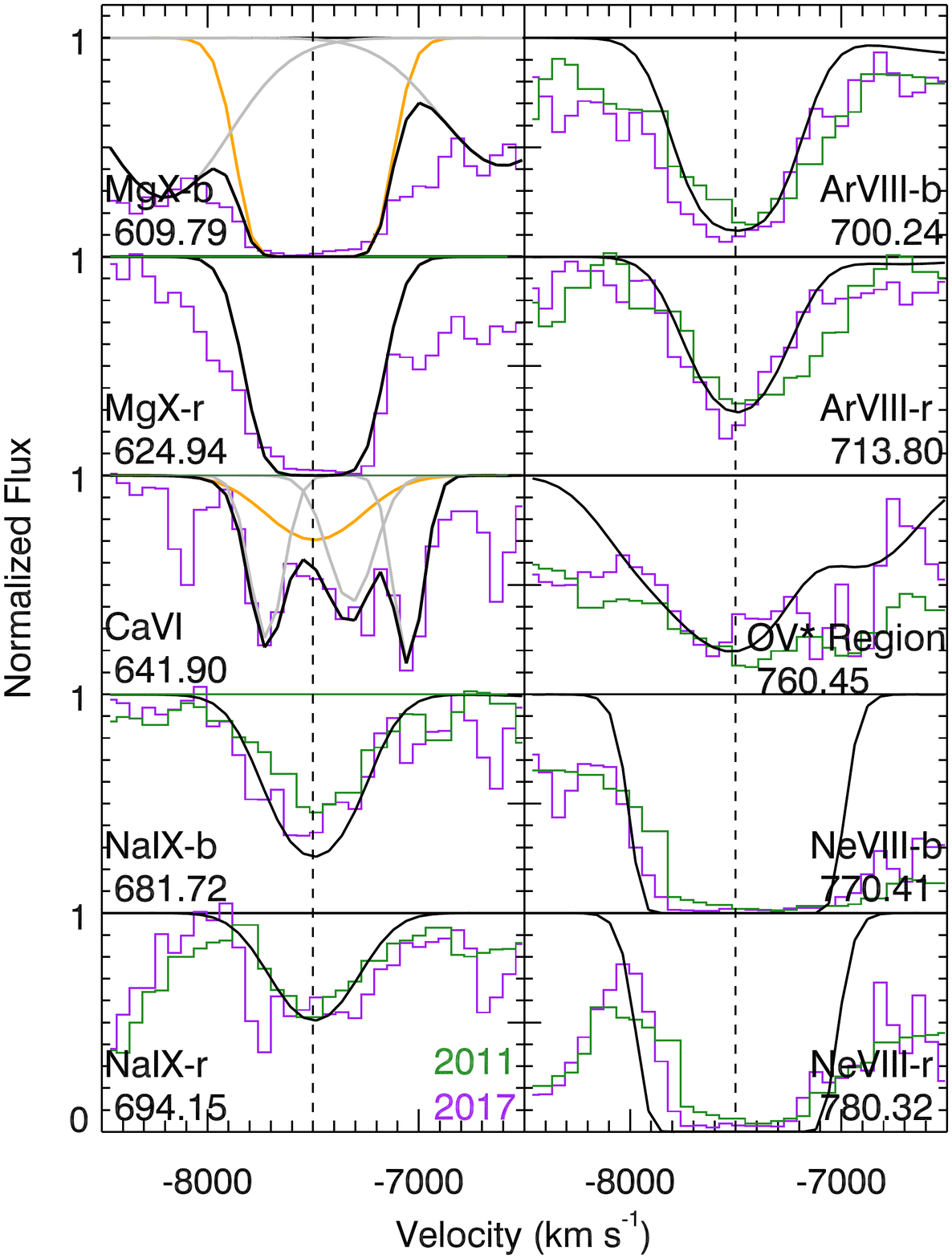}% trim: left lower right upper

\caption{Comparison of data and \SSS\ models for absorption troughs in S2. The velocity center for this \comp\ is marked as the black dashed line at --7500 km s$^{-1}$. The
normalized spectrum is shown in green for the 2011 epoch and purple for the 2017 one. The best-fitting model for S2 is shown in black solid lines. For the \neviii\ and \mgx\ doublets, both the data and model are consistent with significant saturation. For \cavi\ \ly 641.90, we show the model for \cavi\ and the intervening systems as the orange solid and gray dotted lines, respectively. For \mgx\ \ly 609.79, we show the model for \mgx\ as an orange solid line, where contaminating troughs of \oiv\ \ly 608.40 from S1b (the Gaussian centered at $v$ = --6600 km s$^{-1}$) and S2 (the Gaussian centered at $v$ = --8200 km s$^{-1}$) are shown as gray dotted lines. See section \ref{sec:S2:kine} for details.}
\label{fig:vcut_comp2}
\end{figure}

\begin{figure}[htp]

\centering
	\includegraphics[angle=0,trim={0cm 0.00cm 0.3cm 1.5cm},clip=true,width=1\linewidth,keepaspectratio]{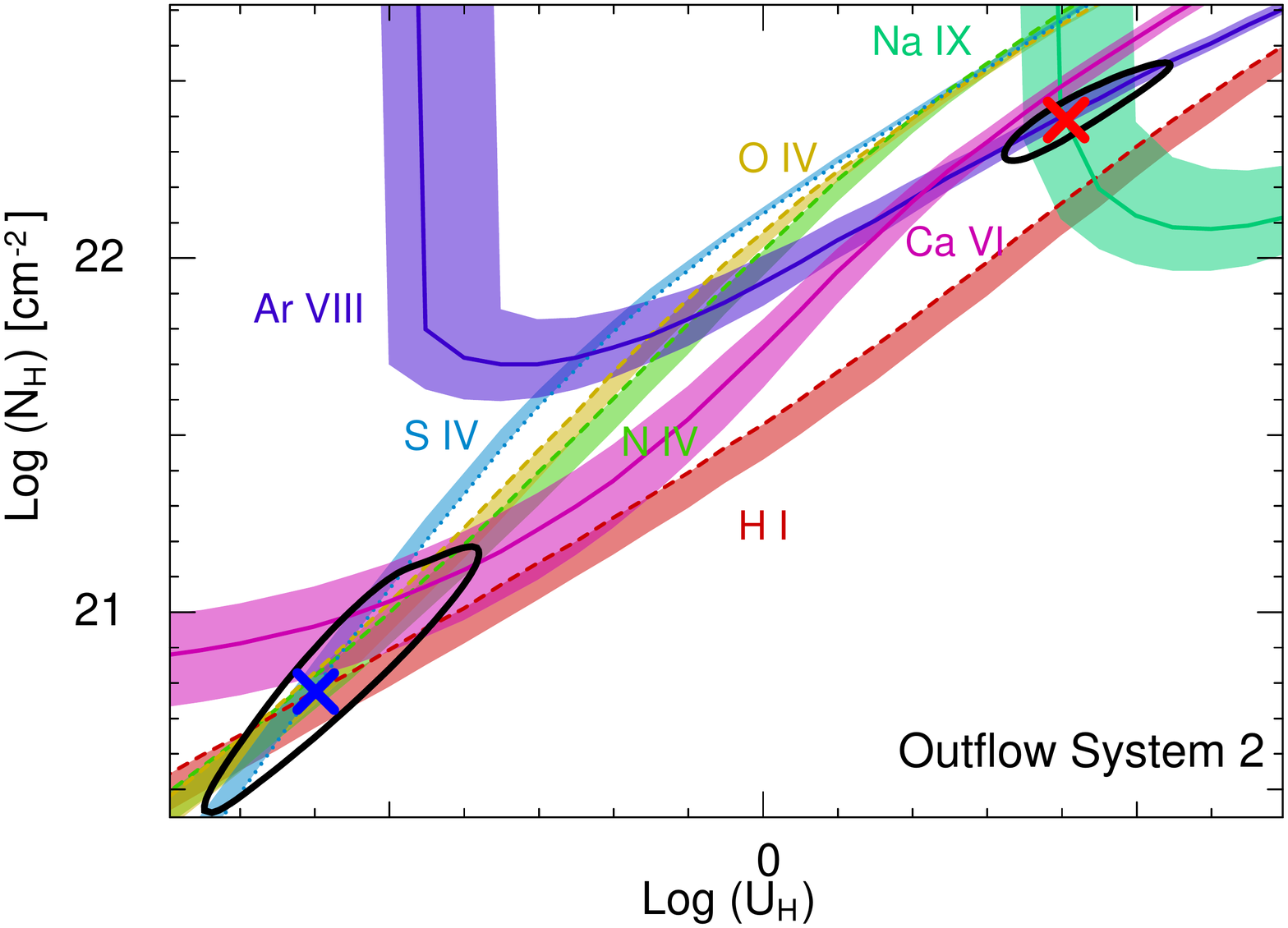} % trim: left lower right upper

\caption{The best-fitting photoionization solution for S2. All patterns and labels are the same as in figure \ref{fig:comp1}. All other N$_{ion}$ in table \ref{tb:IonSystems} that are not shown here are lower or upper limits which are consistent with the solutions and are omitted for clarity's sake.}
\label{fig:comp2}
\end{figure}

We present the best-fitting \SSS\ model as the black solid line in figure \ref{fig:vcut_comp2} for each trough and the corresponding photoionization solution in figure \ref{fig:comp2}. Similar to S1, all of the observed troughs are well fitted by a two-phase photoionization solution. The VHP is constrained mainly by the N$_{ion}$ measurements from \naix\ and \arviii, and is supported by the N$_{ion}$ from \cavi. This yields log(\Nh) = 22.40$^{+0.12}_{-0.10}$ and log(\Uh) = 0.4$^{+0.12}_{-0.06}$. The HP is constrained by the upper limit from \siv, measurement of \cavi, and the lower limits from \oiv\ and \niv. This yields log(\Nh) = 20.79$^{+0.32}_{-0.34}$, and log(\Uh) = --0.60$^{+0.17}_{-0.14}$. Other ions from S2 that are not shown in figure \ref{fig:comp2} are consistent with the solutions. 

\subsubsection{Determination of \ne\ From the \ov*\ Multiplet}
\label{sec:ne_comp2}

\begin{figure}[htp]

\centering
	\includegraphics[angle=0,trim={0.7cm 3.5cm 0.5cm 1.8cm},clip=true,width=1\linewidth,keepaspectratio]{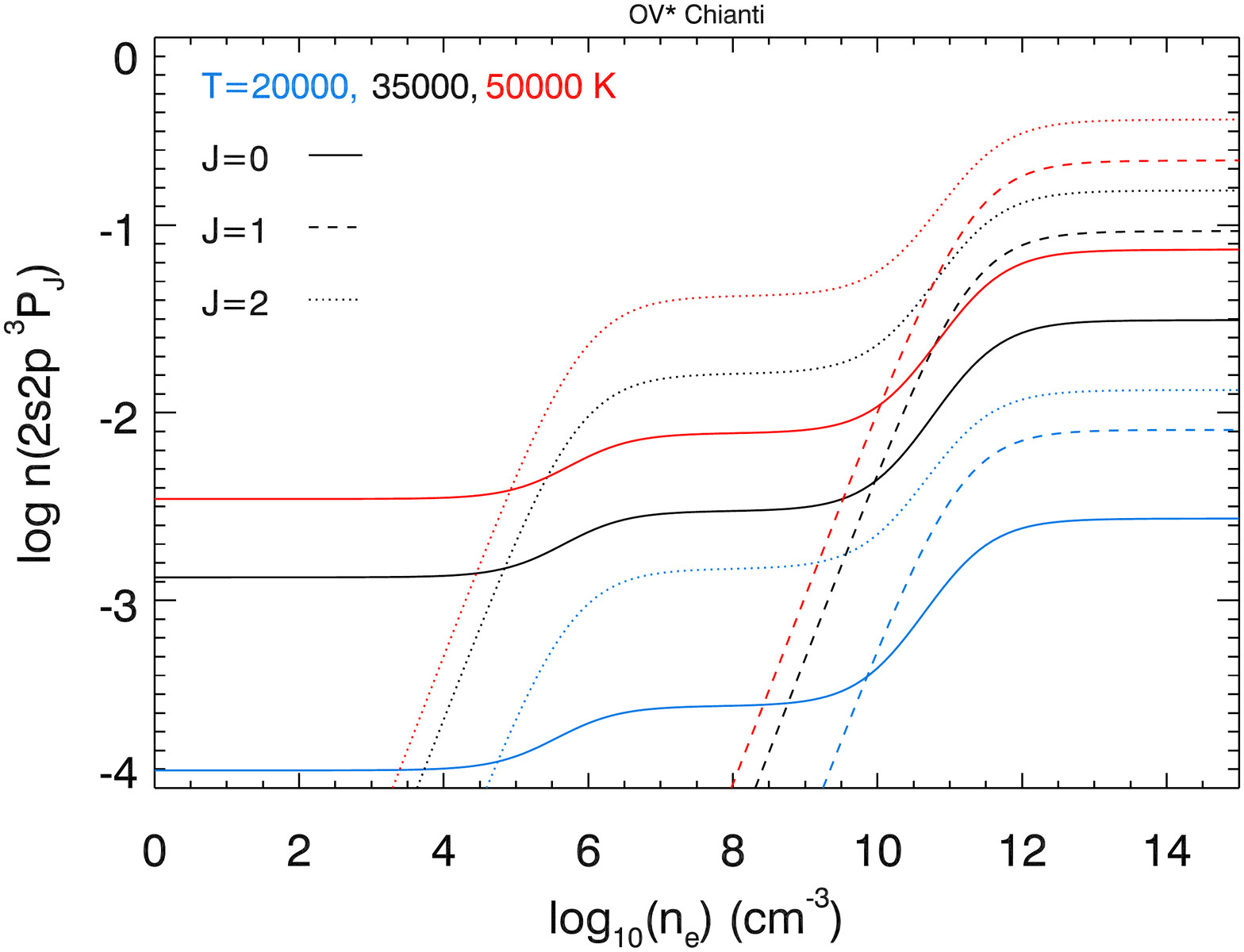}\\% trim: left lower right upper
	\includegraphics[angle=0,trim={0.7cm 0.5cm 0.5cm 1.8cm},clip=true,width=1\linewidth,keepaspectratio]{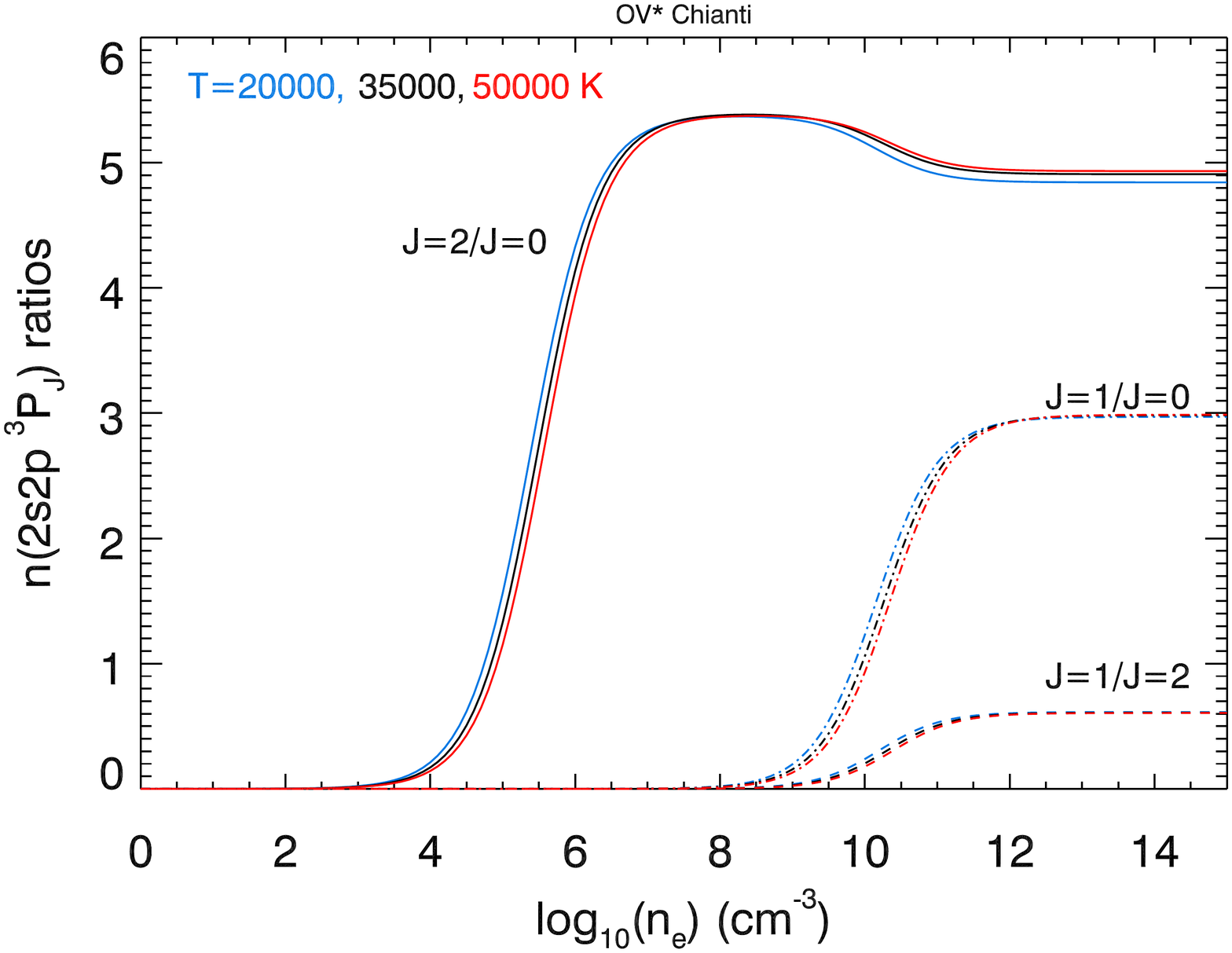}
\caption{\ov*\ level populations. \textbf{Top: }The computed populations for the $^{3}$P$_{J}$ levels of the 2s2p term of O$^{+4}$ are plotted as a function of electron number density [CHIANTI version 7.1.3 \cite[][]{Landi13}]. We show the J = 0, 1, 2 levels in solid, dashed, and dotted lines, respectively. We show the temperature dependence of n(\ov*)/n(\ov) by presenting three different temperature curves. \textbf{Bottom: } Population ratios between different J levels of \ov*. See detailed discussions in section \ref{sec:ne_comp2}.}
\label{fig:ne_comp2_chianti}
\end{figure}

\begin{figure}[htp]

\centering
	\includegraphics[angle=0,trim={0cm 7.5cm 0cm 4cm},clip=true,width=1\linewidth,keepaspectratio]{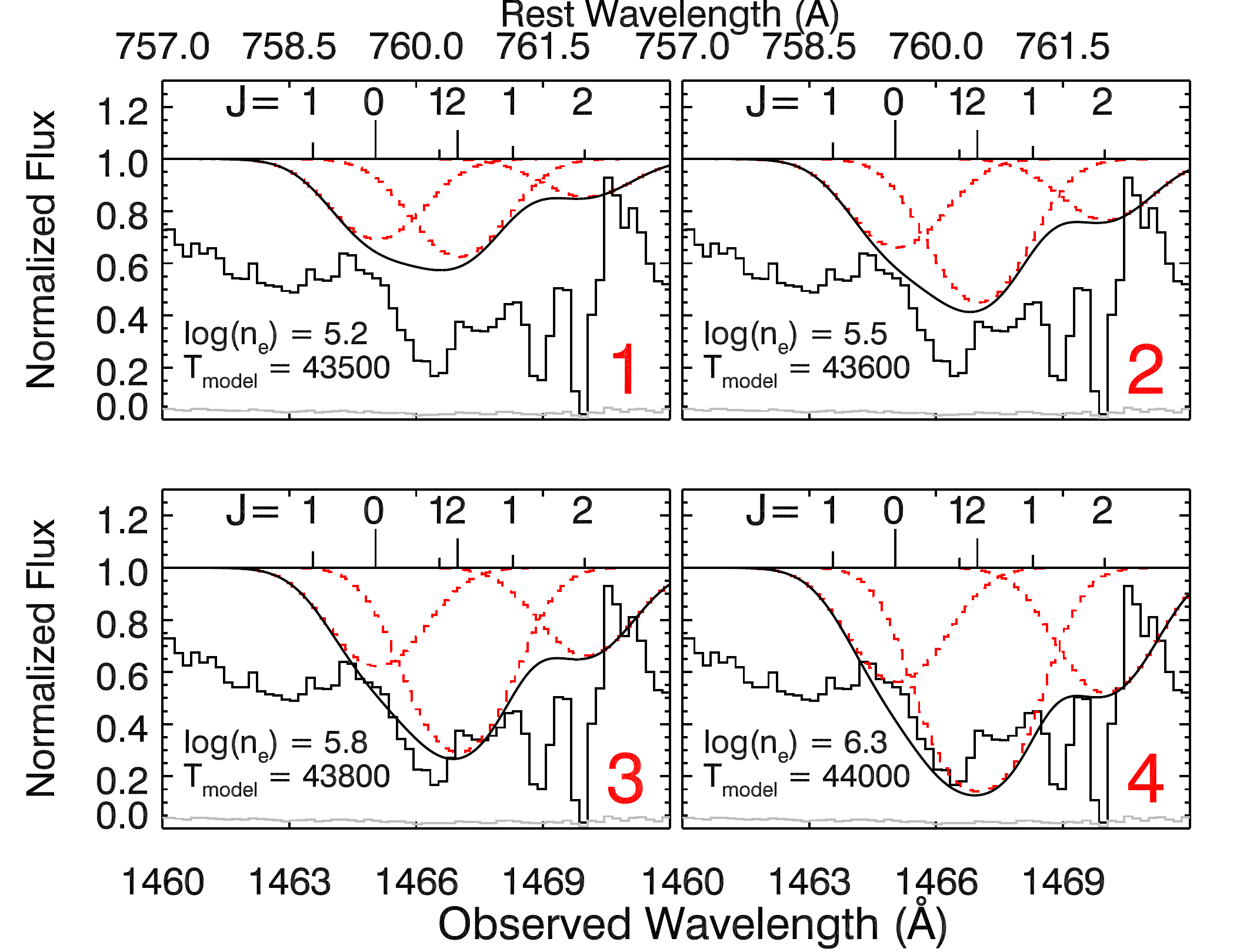}\\% trim: left lower right upper
\caption{Fits to the \ov*\ multiplet region for S2, where we vary \ne\ to get the best fit. The \ne\ of S2 and the corresponding temperature from Cloudy are shown at the bottom-left corner of each panel. The black and gray solid histograms are the normalized flux and errors for the 2017 epoch. For each subplot, the red dashed lines represent the models of the \ov*\ multiplet for a particular \ne, while the black solid lines are the summation of all models in this region. The sharp features at $\sim$1466.5\angstrom, 1469.0\angstrom\  and 1470.0\angstrom\ are intervening systems. We label the four panels as 1 -- 4 at the bottom-right corner. See section \ref{sec:ne_comp2} for a detailed discussion.\\}
%(except \arvi\ since it requires partial covering)
\label{fig:ne_comp2}
\end{figure}

The \ov*\ transitions near 760\angstrom\ are the isoelectronic sequence of the \ciii*\ transitions near 1175\angstrom, which are sensitive to a wide range of \ne\ \cite[][]{Gabel05, Borguet12b, Arav15, Leighly18}. Our HST program shows the first clear detection of \ov*\ absorption troughs in an AGN outflow, in both quasar SDSS J1042+1646 and PKS 0352-0711 (Paper V). For the \ov*\ multiplet, there are six (J -- J$^{'}$) components of the 2s2p$^{3}$P$_{J}$ -- 2p$^{2}$ $^3$P$_{J^{'}}$ multiplet between 758.7\angstrom\ and 762.0\angstrom\ \cite[][]{Doyle83, Keenan94}. However, they are very rare to be observed in either emission or absorption. Previously, the only extra-terrestrial detections of \ov*\ were in solar spectra. The \ov*\ multiplet has been resolved as emission lines by the Harvard S-055 EUV spectrometer onboard Skylab \cite[][]{Doyle83, Keenan94} and more recently by the Solar Ultraviolet Measurements of Emitted Radiation (SUMER) on board the Solar and Heliospheric Observatory (SOHO) \cite[][]{Oshea00}.    

In the upper panel of figure \ref{fig:ne_comp2_chianti}, the ratio of each J level's population to that of the ground state is plotted as a function of \ne. In the bottom panel, we show the population ratios between different J levels versus \ne. When log(\ne) $<$ 5, the level population of \ov*$_{J=0}$, i.e., n(\ov*$_{J=0}$), dominates, and n(\ov*$_{J=1}$) and n(\ov*$_{J=2}$) keep increasing as \ne\ grows in this region. In the log(\ne) range between 5 and 9, n(\ov*$_{J=2}$) $>$ n(\ov*$_{J=0}$) $>$ n(\ov*$_{J=1}$). n(\ov*$_{J=0}$) is populated quickly for high \ne, and finally reachs equilibrium at log(\ne) $>$ 12, where n(\ov*$_{J=2}$) $>$ n(\ov*$_{J=1}$) $>$ n(\ov*$_{J=0}$).

We observed significant absorption troughs at the expected locations of the \ov*\ transitions. The best-fitting photoionzation solution from \SSS\ model predicts that the total log[N(\ov)] = 17.02 cm$^{-2}$, with the ratio of contributions from the HP and VHP as $\sim$ 1:2. The Cloudy predicted temperature for the HP and VHP are around T$_{low}$ = 21,000 K and T$_{high}$ = 44,000 K, respectively. Due to the strong temperature dependence of n(\ov*)/n(\ov), the VHP produces 20 times more N(\ov*)/N(\ov) compared to the HP. Overall, most ($\sim$97.5\%) of the \ov*\ observed in the spectrum is from the VHP.

In order to fit the observed \ov*\ region, we adopt the model predicted value for N(\ov) and temperature T$_{high}$. We vary log(\ne) from 5 to 13, in steps of 0.1 dex and show some of the fits to this range in figure \ref{fig:ne_comp2}. The black and gray histograms are the normalized flux and error for the 2017 epoch, respectively. The red dashed lines are the models for the \ov*\ multiplet with the corresponding J values. The vertical line below each J value indicates the relative $\mathit{f}$ of the line. We note that there are several ``contaminations" from intervening systems at 1466.5\angstrom, 1469.0\angstrom\, and 1470.0\angstrom.
%2) absorption troughs of \arvi\  \ly 754.93 from components 1a and 1b of S1; 3) absorption troughs of the \niv\ \ly 765.15 from S3. 

Despite all of the intervening systems, we can still estimate \ne\ as shown in figure \ref{fig:ne_comp2}. The lower limit is determined at log(\ne) = 5.5, where the model clearly underestimates the data by more than 1$\sigma$ (see panel 1 and 2). The upper limit is determined to be log(\ne) = 6.3, where the model overestimates the data (see panel 4). We get the best fit when log(\ne) = 5.8. Therefore, log(\ne) = 5.8$^{+0.5}_{-0.3}$. 

%The above \ne\ determination is robust. We adopt the predicted value for N(\ov) and temperature from the best-fitting photoionization solution (from section \ref{sec:OS2:PI}). We only vary the \ne\ and are able to fit the observed \ov*\ multiplet absorptions well. It is not presumable to get good fits to the data unless both the photoionization solution and \ne\ are close to accurate.
%In the range of 5.7 $\leq$ log(\ne) $\leq$ 6.5, the overall model gives acceptable fits to the data within the error size.

%3. the absorption troughs of the non-black saturated \niv\ \ly 765 of S3 (see section \ref{sec:comp3}), which is represented as the purple dashed lines. The overall models are showing as the black solid lines and summed all the absorptions in the range except the \niv. This is acceptable as the \niv\ from component 3 is non-black saturated. When adding them together with other absorptions, we need to be careful as the solutions are different if they cover different regions of the cloud (if their covering region overlaps). 

% despite all these contaminations, we can still get the upper and lower bounds of the \ne. We do the experiments to vary log(\ne) of S2 from 8 to 15 by a step size of 0.1 dex. The lower limit of log(\ne) = 11.2, and origins from the discrepancy that the model underestimates the data (see panel 1 and 2). The upper limit of log(\ne) = 11.8 comes from the overestimates of the model to the data (see panel 3 and 4). In the range of 11.2 $\leq$ log(\ne) $\leq$ 11.8, the overall model gives acceptable fits to the data within the error size.\\

Besides \ov*, we observed troughs from other density sensitive transitions. However, they are either situated in a saturated region (\nev*\ \ly 572.33 and \oiv*\ \ly 790.20), or too weak to form a significant trough (\siv*\ \ly 661.40 and the corresponding \siv\ at 657.32\angstrom). From the best-fitting photoionization solution model with the above-determined \ne, we are able to predict absorption troughs that are consistent with the data in the expected regions of these transitions.

\subsection{Outflow \Comp\ 3 (v = --9940 km s$^{-1}$)}
\label{sec:comp3}

\begin{figure}[htp]

\centering
	\includegraphics[angle=0,trim={0cm 0cm 0cm 9.5cm},clip=true,width=1\linewidth,keepaspectratio]{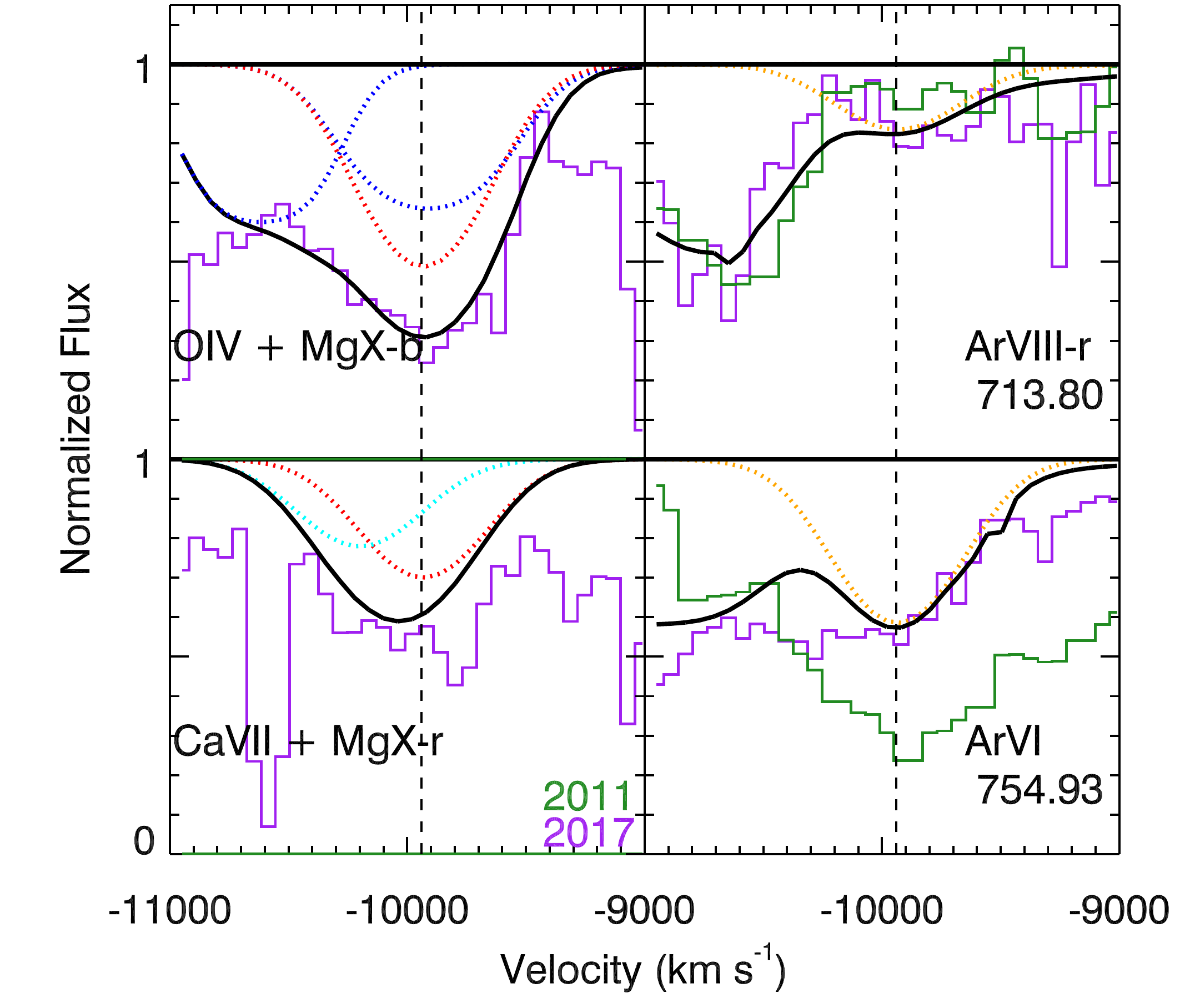}% trim: left lower right upper

\caption{Comparison of the data with the \SSS\ models for the absorption troughs in S3. The velocity center for this \comp\ is the black dashed line at --9940 km s$^{-1}$. The
normalized spectrum is shown in green for the 2011 epoch and purple for the 2017 one. The best-fitting \SSS\ models combining all outflow systems are shown as black solid lines. See section \ref{sec:comp3} for details.}
\label{fig:vcut_comp3}
\end{figure}

\begin{figure}[htp]

\centering
	\includegraphics[angle=0,trim={0cm 0.00cm 0.3cm 1.5cm},clip=true,width=1\linewidth,keepaspectratio]{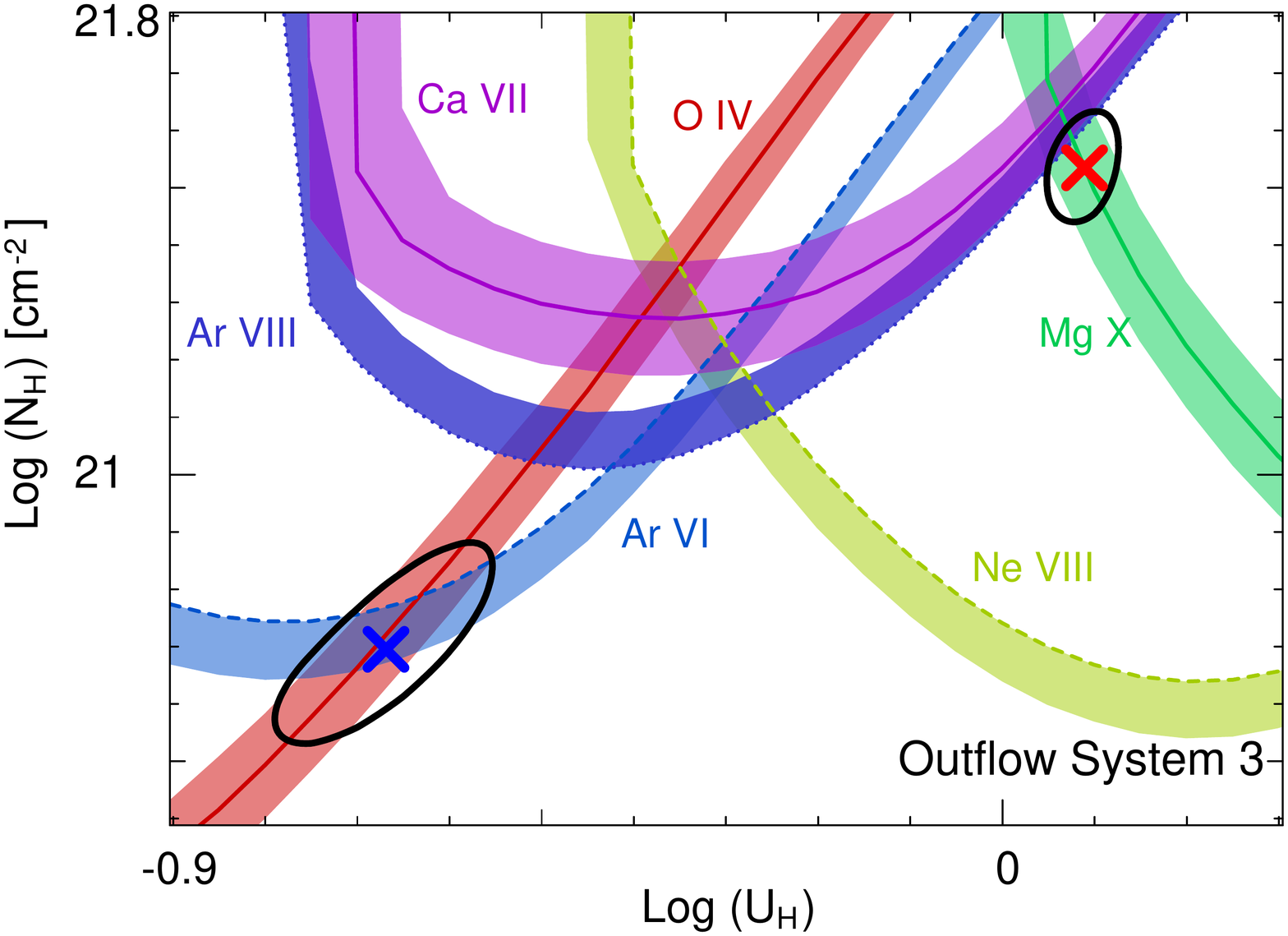} % trim: left lower right upper

\caption{The best-fitting photoionization solution for S3. All patterns and labels are the same as figure \ref{fig:comp1}.}
\label{fig:comp3}
\end{figure}

S3 also shows absorption troughs from both high- and very high-ionization species. We show the data and the best-fitting \SSS\ models in figure \ref{fig:vcut_comp3}. For the \oiv\ + \mgx-b region, the blue dotted lines are for the \oiv\ \ly 608.340 and \oiv*\ \ly 609.83 troughs while the red dotted line is for the \mgx\ \ly 609.79 trough. For the \cavii\ + \mgx-r region, the cyan and red dotted lines are for the \cavii\ \ly 624.28 and \mgx\ \ly 624.94 troughs, respectively. These two regions fall in the gap of the HST/COS G140L grating in the 2011 epoch. For the \arviii-r and \arvi\ region, their modeled troughs are shown as orange dotted lines. The \arviii-r trough is consistent with the shallow observed absorption and is contaminated by the \arviii\ \ly 700.24 trough from S1a (centered at -10,600 km s$^{-1}$ in the top-right panel of figure \ref{fig:vcut_comp3}). Therefore, the AOD \Nion\ from the \arviii-r trough is treated as an upper limit. The higher velocity wing of the \arvi\ \ly 754.93 trough is contaminated by the \neviii\ \ly 780.32 trough from S4, while S4 shows variability between the 2011 and 2017 epochs. We discuss this varibility in Paper IV.

We report the measured \Nion\ in table \ref{tb:IonSystems2} and the best-fitting photoinoization solution from the \SSS\ model in figure \ref{fig:comp3}. All observed outflow absorption troughs are well fitted by a two-phase ionization model. The observed amount N(\cavii) originates from both the HP and VHP, with a ratio of $\sim$ 1:3, respectively. Therefore, \cavii\ gives constraints to both phases. The HP yields log(\Nh) = 20.69$^{+0.34}_{-0.28}$ and log(\Uh) = --0.67$^{+0.21}_{-0.31}$. This solution is constrained by the \oiv\ and \arvi\ troughs, as well as the \cavii\ trough. The VHP gives log(\Nh ) = 21.54$^{+0.17}_{-0.18}$ and log(\Uh) = 0.09$^{+0.08}_{-0.07}$. This solution is constrained mainly by the measured N$_{ion}$ of the \mgx\ doublet and \cavii, and the upper limit of \arviii.  Comparing between the 2011 and 2017 epochs, the troughs of S3 show no variations.

We do not observe measurable absorption troughs from density diagnostic transitions for S3. Therefore, the \ne\ of S3 is undetermined.

\begin{turnpage}
\begin{deluxetable*}{c c c c c c c c c c c  l}[htb!]
\tablewidth{1.0\textwidth}
\tabletypesize{\small}
\tablecaption{Analysis Results For the Outflows Seen in SDSS J1042+1646\label{table:compareTable}}
\tablehead{
 \colhead{System}   & \colhead{ V} & \colhead{log(\Uhhp)} & \colhead{log(\Nhhp)} & \colhead{log(\ne)}& \colhead{log(\Uhvhp)} & \colhead{log(\Nhvhp)} & \colhead{log($\vy{f}{V}$)$^{(a)}$}& \colhead{R} & \colhead{$\dot{M}$}& \colhead{Log $\dot{E_{k}}$} & \colhead{ $\dot{E_{k}}/L_{Edd}$}
\\ [-2mm]
\\
 \colhead{}   & \colhead{(km s$^{-1}$)} & \colhead{log}& \colhead{log(cm$^{-2}$)} & \colhead{log(cm$^{-3}$)}& \colhead{log}  & \colhead{log(cm$^{-2}$)} &\colhead{}	   & \colhead{pc} & \colhead{($M_{\odot}$ yr$^{-1}$)} & \colhead{log(erg s$^{-1}$)} & \colhead{$\%$} 
}

%\toprule
%\specialrule{0.01em}{0.2em}{0.6em}
\startdata

\multicolumn{11}{l}{\textbf{HE0238 SED:}}\\ 	
\hline	
\textbf{1a} 	&\textbf{-4950}	&\textbf{-1.0$^{+0.2}_{-0.3}$}&\textbf{20.4$^{+0.4}_{-0.6}$}&\textbf{3.7$^{+0.2}_{-0.3}$} 	&\textbf{0.4$^{+0.2}_{-0.1}$}&\textbf{22.4$^{+0.2}_{-0.1}$}	& \textbf{-3.4$^{+0.5}_{-0.7}$}
&\textbf{840$^{+500}_{-300}$}&\textbf{2800$^{+200}_{-800}$} &\textbf{46.4$^{+0.1}_{-0.1}$}&\textbf{10$^{+3}_{-2}$}\\

\textbf{1b} 	&\textbf{-5750}	&\textbf{-0.9$^{+0.2}_{-0.2}$}&\textbf{20.5$^{+0.4}_{-0.3}$}&\textbf{3.8$^{+0.2}_{-0.3}$} 	&\textbf{0.5$^{+0.2}_{-0.2}$}&\textbf{22.5$^{+0.3}_{-0.2}$}	& \textbf{-3.4$^{+0.6}_{-0.5}$}	
&\textbf{800$^{+300}_{-200}$}&\textbf{4300$^{+1200}_{-1500}$} &\textbf{46.7$^{+0.2}_{-0.1}$}&\textbf{20$^{+14}_{-4}$}\\

\textbf{2} 	&\textbf{-7500}	&\textbf{-0.6$^{+0.2}_{-0.1}$}&\textbf{20.8$^{+0.3}_{-0.3}$}&\textbf{5.8$^{+0.5}_{-0.3}$}	&\textbf{0.4$^{+0.1}_{-0.1}$}&\textbf{22.4$^{+0.1}_{-0.1}$}	& \textbf{-2.6$^{+0.4}_{-0.6}$}
&\textbf{15$^{+8}_{-8}$}&\textbf{81$^{+20}_{-30}$} &\textbf{45.1$^{+0.1}_{-0.2}$}&\textbf{0.5$^{+0.2}_{-0.2}$}\\

\textbf{3} 	&\textbf{-9940}	&\textbf{-0.7$^{+0.1}_{-0.1}$}&\textbf{20.7$^{+0.2}_{-0.2}$}&\textbf{--}			&\textbf{0.1$^{+0.1}_{-0.1}$}&\textbf{21.5$^{+0.1}_{-0.1}$}	& \textbf{-1.6$^{+0.6}_{-0.3}$}	
&\textbf{--}&\textbf{--} &\textbf{--}&\textbf{--}\\

%\textbf{OS 4} 	&\textbf{-21050}&\textbf{--}			 &\textbf{--}			  &\textbf{--}		&\textbf{0.14$^{+0.07}_{-0.08}$}	&\textbf{20.95$^{+0.13}_{-0.13}$}	
%&\textbf{--}	&\textbf{--}&\textbf{--}&\textbf{--}	 		 \\
\hline
\\ [+0.01mm]
\multicolumn{11}{l}{\textbf{Transmitted HE0238 SED Emerging From the VHP of OS 2$^{(b)}$:}}\\
\hline
\textbf{1a} 	&\textbf{-4950}	&\textbf{--0.7$^{+0.2}_{-0.2}$}&\textbf{20.6$^{+0.3}_{-0.3}$}&\textbf{3.7$^{+0.2}_{-0.3}$} 	&\textbf{1.0$^{+0.2}_{-0.1}$}&\textbf{22.5$^{+0.3}_{-0.2}$}	& \textbf{-3.6$^{+0.4}_{-0.4}$}	
&\textbf{600$^{+200}_{-200}$}&\textbf{2700$^{+500}_{-100}$} &\textbf{46.4$^{+0.1}_{-0.1}$}&\textbf{8$^{+2}_{-2}$}\\

\textbf{1b} 	&\textbf{-5750}	&\textbf{--0.6$^{+0.2}_{-0.2}$}&\textbf{20.6$^{+0.3}_{-0.3}$}&\textbf{3.8$^{+0.2}_{-0.3}$} 	&\textbf{1.1$^{+0.3}_{-0.2}$}&\textbf{22.6$^{+0.4}_{-0.3}$}	& \textbf{-3.7$^{+0.5}_{-0.6}$}	
&\textbf{530$^{+200}_{-200}$}&\textbf{3500$^{+2500}_{-700}$} &\textbf{46.6$^{+0.2}_{-0.1}$}&\textbf{16$^{+8}_{-3}$}\\

\textbf{3} 	&\textbf{-9940}	&\textbf{--0.2$^{+0.1}_{-0.1}$}&\textbf{21.2$^{+0.1}_{-0.1}$}&\textbf{--}			&\textbf{1.1$^{+0.1}_{-0.3}$}&\textbf{20.9$^{+0.4}_{-0.1}$}	& \textbf{-1.0$^{+0.3}_{-0.4}$}
&\textbf{--}&\textbf{--} &\textbf{--}&\textbf{--}\\
\hline
\\ [+0.01mm]
\multicolumn{11}{l}{\textbf{Comparison to Other Energetic Outflows:$^{(c)}$}}\\
\hline
\textbf{HE0238--1904} 	&\textbf{-5000}	&\textbf{-1.5$^{+0.6}_{-0.7}$}&\textbf{17.6$^{+0.5}_{-0.1}$}&\textbf{3.7$^{+0.1}_{-0.1}$} 	&\textbf{0.4$^{+0.1}_{-0.1}$}&\textbf{19.6$^{+0.2}_{-0.1}$}& \textbf{-3.9$^{+0.8}_{-0.7}$}	
&\textbf{3400$^{+900}_{-2800}$}&\textbf{160$^{+80}_{-150}$} &\textbf{45.7$^{+0.2}_{-1.2}$}&\textbf{4$^{+2}_{-3}$}\\

\textbf{J0831+0354} 	&\textbf{-10800}&\textbf{-0.2$^{+0.4}_{-0.5}$}&\textbf{22.4$^{+0.5}_{-0.5}$}&\textbf{4.4$^{+0.3}_{-0.2}$} 	&\textbf{--}&\textbf{--}				& --
&\textbf{80$^{+27}_{-18}$}&\textbf{230$^{+330}_{-130}$} &\textbf{45.9$^{+0.4}_{-0.3}$}&\textbf{8$^{+11}_{-4}$}\\

\vspace{-2.2mm}
\enddata

\tablecomments{The bolometric luminosity of SDSS J1042+1646 is $\sim$ 1.5 $\times$ 10$^{47}$ erg s$^{-1}$.\\
%(1). Table of the best-fitting results from the \SSS\ models and derived outflow parameters for each outflow \comp.\\
(a). The volume filling factor of the outflow's high-ionization phase to the very high-ionization phase (see definition in section \ref{sec:2Phases:Factor}).\\
(b). The effect of the transmitted SED is described in section \ref{sec:shading}.\\
(c). The results of outflows from HE0238--1904 and SDSS J0831+0354 are based on the HE0238 SED with solar metallicity \cite[][]{Arav13,Chamberlain15b}.\\
%$^{(a)}$: The fluxes of the two epochs are 3 and 4, which is measured at observed wavelength of 1350\angstrom, in units of 10$^{-15}$ erg s$^{-1}$ cm$^{-2}$\angstrom$^{-1}$
%(3). log(L$_{edd}$) = 47.61 [log(erg s$^{-1}$)], which is estimated following the mass-luminosity relationship found in function (9) of \cite{Peterson04}. 
% I am using the flux = 46E-17 erg/s/cm^2/A at restwave =5100A from the extension of SDSS data (https://dr12.sdss.org/spectrumDetail?mjd=53851&fiber=376&plateid=2480). 
%log(L$_{edd}$) = 47.11 [log(erg s$^{-1}$)], which is estimated following the function (3) in \cite{Park13}.  
\\
}
\label{tb:ParaSystems}
\end{deluxetable*}
\end{turnpage}

\section{Distances and Energetics}
\label{sec:energy}
Using the \ne\ determination for each outflow system, we can solve equation (\ref{Eq:ionPoten}) for $R$. We then determine the mass flow rate ($\dot{M}$) and the kinetic luminosity ($\dot{E_{k}}$) by assuming the outflow is in the form of a thin and partially filled shell \cite[see elaboration in][]{Borguet12a}:
\begin{equation}\label{eq:1}
\begin{split}
\dot{M}\simeq 4\pi \Omega R\Nh \mu m_p v 
\end{split}
\end{equation}

\begin{equation}\label{eq:2}
\begin{split}
\dot{E}_{k}\simeq \frac{1}{2} \dot{M}v^2
\end{split}
\end{equation}

where $\Omega$ is the global covering factor, i.e., the portion of the full solid angle covered by the outflow (for SDSS J1042+1646, we use $\Omega$ = 0.2, see discussion in section \ref{sec:dis1}), $\mu$ = 1.4 is the mean atomic mass per proton, $m_p$ is the proton mass, and $\mathit{v}$ is the velocity of the outflow.

Since the \mgii\ emission region for SDSS J1042+1646 is observed by SDSS, we follow the \mgii--based blackhole mass estimators \cite[calibrated with H$\beta$ reverberation measurement, see sequation (7) and table (4) in][]{Bahk19} to derive a black hole mass of 10$^{9.3}$ and corresponding Eddington luminosity, L$_{\text{Edd}}$ = 10$^{47.4}$ erg s$^{-1}$. We then calculate the ratio of the outflow's kinetic luminosity to the L$_{\text{Edd}}$, i.e., $\Gamma_{\text{Edd}}$ $\equiv$ $\dot{E_{k}}/L_{\text{Edd}}$. 

In table \ref{tb:ParaSystems}, we show all the physical parameters for the outflows we extracted from the data. In the first part of the table \ref{tb:ParaSystems}, we show the derived photoionization solutions, \ne, $\vy{f}{V}$ (volume filling factor, see section \ref{sec:2Phases:Factor}), and the energetics for the four outflows analyzed in section \ref{text:AllComps}. We show the ``shading effect" results in the second part of the table (see section \ref{sec:shading}). Finally, we compare our derived physical parameters to two other energetic outflows in the third part of the table (see discussion in section \ref{sec:dis1}).

\section{Discussion}
\label{sec:discussion}
%\subsection{Outflow Velocity Shifts}
%NGC3783 and NGC7469

\subsection{The Two-phase Outflows}
\label{sec:2Phases}
\subsubsection{Prevalence of Outflows with a VHP}
Two ionization phase outflows have been reported and discussed in detail for quasar HE0238--1904 \cite[see section 8 in][]{Arav13}. Their lower ionization phase (similar to HP in this paper) was mainly constrained by the \Nion\ from \hi\ and \oiv, while the higher-ionization phase (similar to VHP in this paper) was constrained by the \Nion\ from \ovi, \mgx, and \neviii. Similarly, all of the outflows analyzed here (S1a, S1b, S2, and S3) as well as the 8 \XUV\ outflows reported in three additional quasars (see Paper I), necessitate an HP and a VHP. All of these outflows show \oiv\ troughs and therefore should have \civ\ 1549 \AA\ troughs. Thus, these outflows would be classified as HiBALs if observed in rest-frame wavelengths, $\lambda_{\text{rest}}$ $>$ 1050\angstrom\ (see section 4.5 in Paper I). The existence of a VHP in all outflows from these quasars suggests that all HiBALs have VHPs, which is not practicably accessible with ground-based telescopes (see section \ref{Introduction}). The selection criteria of the \XUV\ sample do not affect the extrapolation of these results to the general HiBAL outflow population (see section 4.1 in Paper 1). As shown in table 1 of Paper I (for all 13 outflows described in Paper II -- VI), the VHP has 6 -- 100 times higher \Nh\ than the HP, which indicates that the VHP also carries the bulk of the material in the outflows.

\subsubsection{Volume Filling Factor}
\label{sec:2Phases:Factor}
Kinematic similarities (both velocity and width centroid) between troughs from the HP and the VHP suggest that these two phases are co-spatial. For each phase (HP or VHP), the volume is proportional to \Nh/$\vy{n}{H}$, and the $\vy{n}{H}$ ratio between the HP and the VHP is given by $U_{\text{H},\scriptscriptstyle{VHP}}/U_{\text{H},\scriptscriptstyle{HP}}$. Therefore, the volume filling factor between the two phases is defined as \cite[see section 8.1 in][]{Arav13}:

\begin{equation}\label{eq:Vfactor}
\begin{split}
\vy{f}{V} \equiv\ \frac{V_{\scriptscriptstyle{HP}}}{V_{\scriptscriptstyle{VHP}}} = \frac{N_{\text{H},\scriptscriptstyle{HP}}}{N_{\text{H},\scriptscriptstyle{VHP}}} \times \frac{U_{\text{H},\scriptscriptstyle{HP}}}{U_{\text{H},\scriptscriptstyle{VHP}}}
\end{split}
\end{equation}
%\frac{m_{\scriptscriptstyle{HP}}/m_{\scriptscriptstyle{VHP}}}{\rho_{\scriptscriptstyle{HP}}/\rho_{\scriptscriptstyle{VHP}}}

where $V$ is the volume of the outflows, and the subscripts HP and VHP denotes the high-ionization phase and very high-ionization phase, respectively. For the outflows in quasar HE0238--1904, the VHP has log(\Uh) 2 -- 3 dexes higher and log(\Nh) 2 -- 2.5 dexes higher than the HP. Therefore, the HP is inferred to have log ($\vy{f}{V}$) $\sim$ --4 to --6 compared to the VHP. Thus, \cite{Arav13} reached the conclusion that the VHP carries much more material than the HP and is closer to the situation seen in X-ray warm absorbers \cite[e.g.,][]{Netzer03, Gabel05}. We show the $\vy{f}{V}$ values for the outflows in SDSS J1042+1646 in table \ref{tb:ParaSystems}.

%For our outflows S1 and S2, the VHP has log(\Uh) 1 -- 1.4 dexes higher and log(\Nh) 1.6 -- 2 dexes higher than the HP. Therefore, we have that log($\vy{f}{V}$) $\sim$ --3.4$^{+0.8}_{-1.3}$ and --2.6$^{+0.7}_{-0.6}$, respectively. In outflow S3, the best-fitting solutions have closer log(\Nh) in HP and VHP (see table \ref{table:compareTable}), which yield a log($\vy{f}{V}$) = --1.6$^{+0.8}_{-0.9}$. Comparison of the 3 outflows we analyzed here and those observed in HE0238--1904 suggests that 1) 0.003 $\lesssim$ $\frac{N_{\text{H},\scriptscriptstyle{HP}}}{N_{\text{H},\scriptscriptstyle{VHP}}}$ $\lesssim$ 0.14; 2) 0.001 $\lesssim$ $\frac{U_{\text{H},\scriptscriptstyle{HP}}}{U_{\text{H},\scriptscriptstyle{VHP}}}$ $\lesssim$ 0.17. We compare the physical parameters of our outflows to HE0238--1904 in table \ref{table:compareTable}.

\subsubsection{Comparisons with Warm Absorbers}
In our case, two ionization-phase solutions are sufficient for fitting all the observed outflow features, but it does not exclude more ionization phases. As shown in X-ray warm absorber studies, AGN outflows can span up to 5 orders of magnitude in ionization parameter, i.e., --1 $<$ log($\xi$) $<$ 4 \cite[][]{Steenbrugge03, Costantini07, Holczer07, McKernan07, Behar09}, where $\xi$ is the X-ray ionization parameter and for the HE0238 SED, log(\Uh) = log($\xi$) -- 1.3. Moreover, studies in X-ray invoked a continuous distribution of \Nh\ as a function of $\xi$ \cite[][]{Holczer07}. The observed absorption features in SDSS J1042+1646 are well fitted with just two ionization phases. However, future sensitive X-ray spectroscopy with observatories such as {\it Athena} \cite[][]{Barcons17} may reveal additional higher-ionization phases and/or the necessity of a continuous distribution of \Nh\ as a function of \Uh.

%Therefore, this gives the evidence that we can observe outflows with similar log(\Nh) in the two phases, and the HP can fill comparably more space than the formally report values.
%log($\vy{f}{V}$) $\sim$ --2.6$^{+0.7}_{-0.6}$ -- --3.4$^{+0.8}_{-1.3}$
\subsection{``Shading Effect" of Different Outflow \Comps}
\label{sec:shading}

\begin{figure}[htp]

\centering
	\includegraphics[angle=0,trim={0.35cm 2.45cm 0.0cm 0cm},clip,width=1.0\linewidth,keepaspectratio]{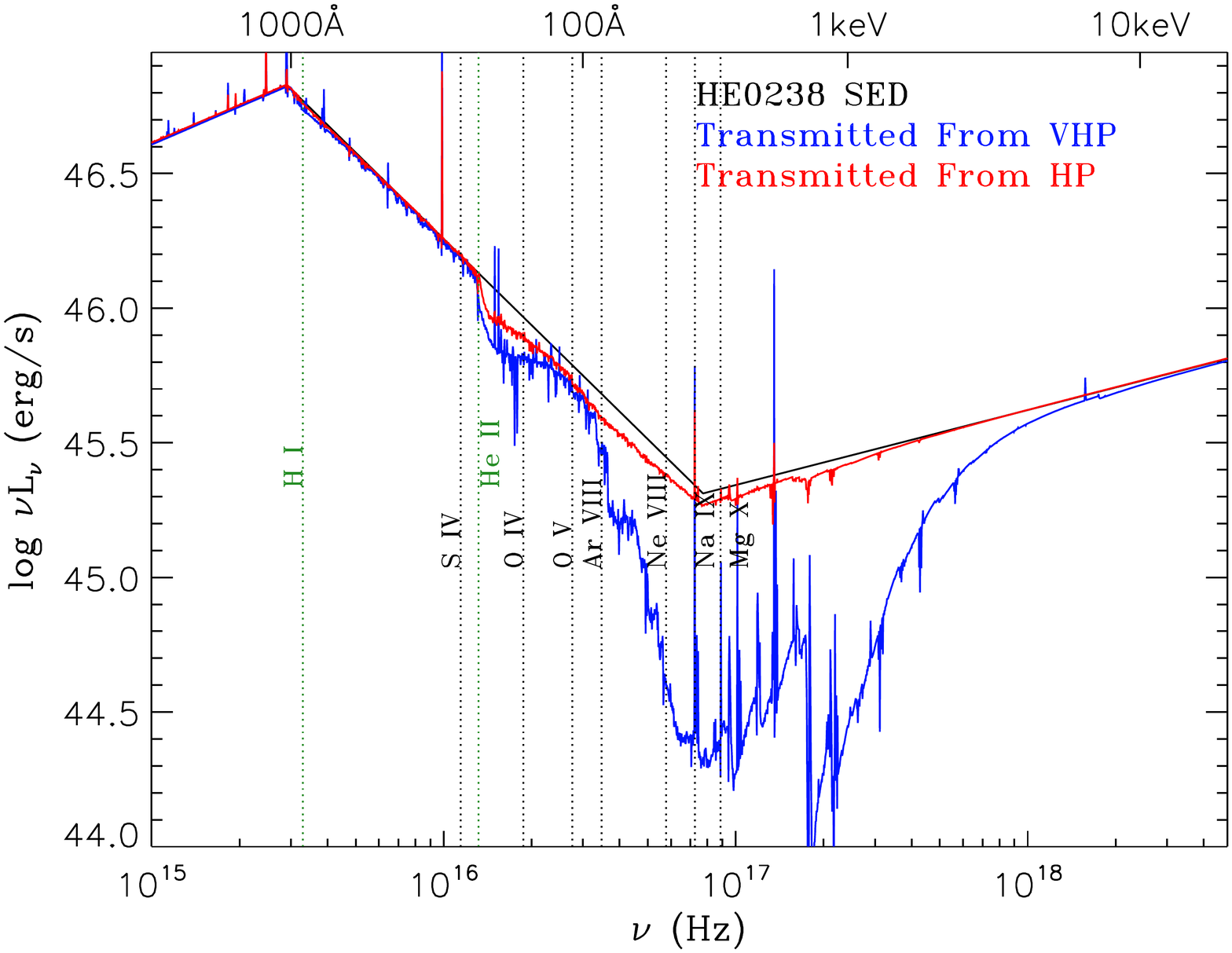}\\ % trim: left lower right upper
	\includegraphics[angle=0,trim={0.35cm 0cm 0.0cm 1.85cm},clip,width=1.0\linewidth,keepaspectratio]{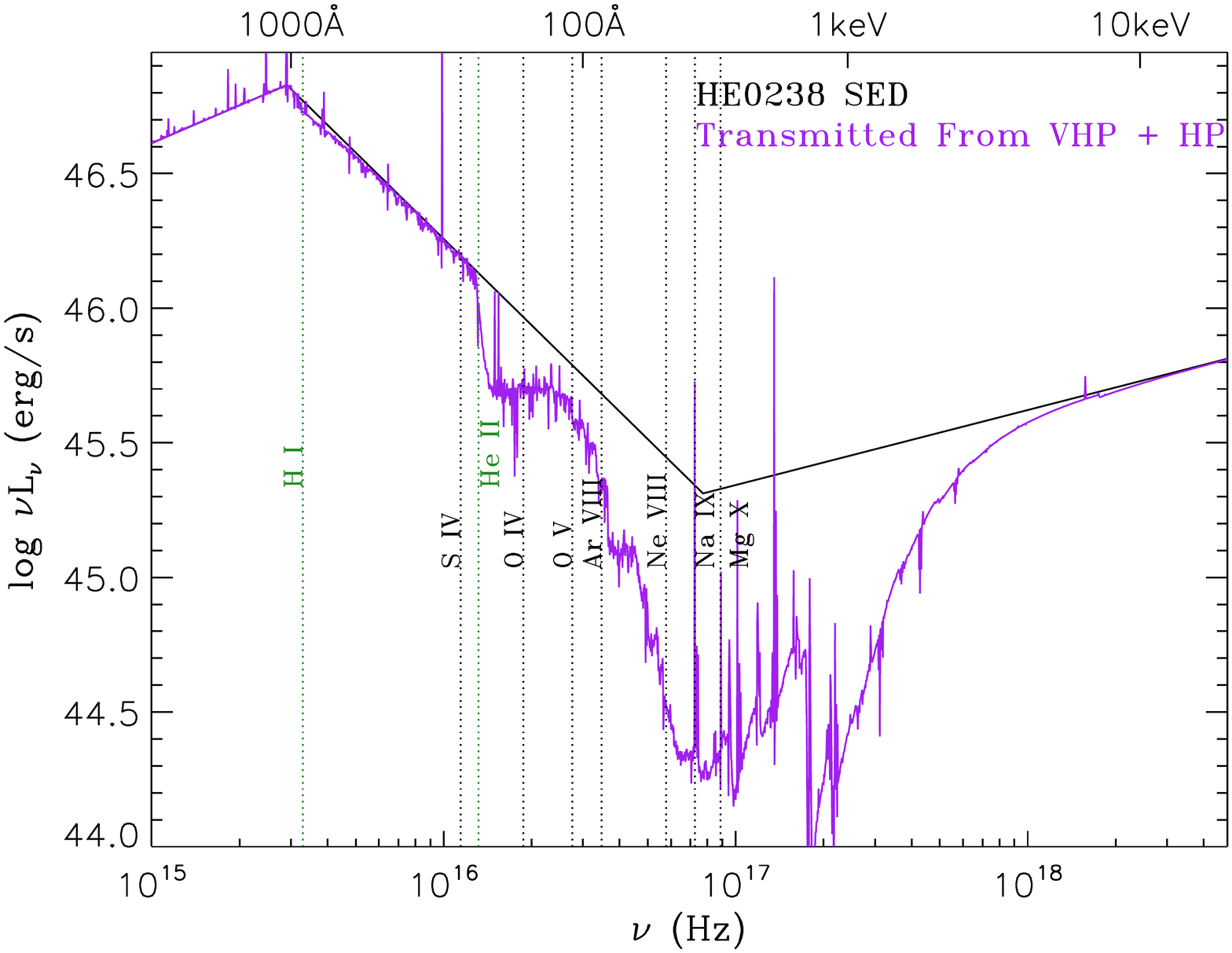}
  
\caption{Comparison of the transmitted SEDs (in blue solid lines) to HE0238 SED (in black solid lines). \textbf{Top:} We show the transmitted SED emerging from the very high- and high-ionization phase of outflow S2 as blue and red lines, respectively. \textbf{Bottom:} We show the transmitted SED emerging from the high-ionization phase of outflow S2 when it is illuminated by SED$_{\text{Tran.VHP}}$. In both plots, we indicate the location of the ionization potentials of the main species by the vertical dotted lines. See detailed discussions in section \ref{sec:shading}.}
\label{fig:Shading}
\end{figure}

Since we have multiple outflow systems along the line of sight at different distances, the inner outflows would absorb photons from the central AGN and could change the SED incident on the outer outflows. This is called the ``Shading Effect" \cite[][]{Sun17} and here we quantify its effect on our derived photoionization solutions. We show our results when the outflow is directly exposed to the HE0238 SED in the first part of table \ref{tb:ParaSystems}. Outflow S2 has a smaller $R$ compared to the other \comps, and its high \Nh\ could significantly affect the SED seen by other outflows. Therefore, we test the shading covered by S2 here and compare it to the original results using the unshaded HE0238 SED.

Our model predicts that S2 has two ionization phases. We input the HE0238 SED and calculate the transmitted SED after passing through S2's VHP and HP to get SED$_{\text{Tran.VHP}}$ and SED$_{\text{Tran.HP}}$, respectively. We show these two transmitted SEDs in the top panel of figure \ref{fig:Shading} as blue and red lines, respectively. The unshaded HE0238 SED is plotted as the solid black line. The SED$_{\text{Tran.VHP}}$ shows a reduction of flux $\sim$ 0.3 dex near the \heii\ edge; and further reduction of flux $\sim$ 1 dex from 100 eV to 1 keV, which is mainly caused by metal absorptions. The SED$_{\text{Tran.HP}}$ has a minimal reduction while the largest change is $\sim$ 0.1 dex near the \heii\ edge. To check the total ``Shading Effect", we test the scenario where SED$_{\text{Tran.VHP}}$ is incident on the HP of S2. As shown in the bottom panel of figure \ref{fig:Shading}, considering the shading from both phases (purple lines) only affects the SED by an additional 10\% compared to the SED$_{\text{Tran.VHP}}$ (blue lines in the top panel). 
%Therefore, the ``Shading Effect" of the VHP is the main contributor.
%The total ``Shading Effect" realistically needs to consider the spatial distribution of HP and VHP, which is unknown here.

We use these transmitted SEDs as the input SEDs to generate new grids of models with Cloudy. Based on these models and the same method reported in section \ref{text:AllComps}, we find new photoionization solutions for the shaded \comps\ S1a, S1b, and S3.  We then calculate new Q$_{H}$ values using the transmitted SEDs and calculate the outflows' R, and energetics accordingly [see equations (\ref{Eq:ionPoten}), (\ref{eq:1}), (\ref{eq:2})].  We compare these solutions to the un-shaded results in table \ref{tb:ParaSystems}, where the second part of the table is for the photoionization solutions from SED$_{\text{Tran.VHP}}$. We find that when considering the ``Shading Effect" with the SED$_{\text{Tran.VHP}}$, we get a larger log(\Uh) by $\sim$ 0.3  -- 1.0 dex and a larger log(\Nh) by $\sim$ 0.1  -- 0.5 dex. This is because the shading of S2 decreases the amount of hydrogen ionizing photons, so a higher \Uh\ is needed to keep a similar ionization state as the unshaded case. The new \Uh\ and \Nh\ values lead to a decrease of $\sim$ 0.1 dex in R, $\dot{M}$ and $\dot{E_{k}}$. 

\subsection{AGN Feedback Effects and Global Covering Factor}
\label{sec:dis1}
As shown in section \ref{Introduction}, significant AGN feedback typically requires high $\dot{E_{k}}$, where $\Gamma_{\text{Edd}}$ is at least 0.5 \cite[][]{Hopkins10} or 5\% \cite[][]{Scannapieco04}. Equations (\ref{eq:1}) and (\ref{eq:2}) show that $\dot{M}$ and $\dot{E_{k}}$ are linearly dependent on the solid angle subtended by the outflows around the source (4$\pi\Omega$). Since there are no direct spectroscopic determinations for $\Omega$, the common approach is to use the detection frequency of outflows in surveys as a proxy. In SDSS J1042+1646, the most energetic outflow is S1b. Based on our analysis results, $\dot{M}$ and $\dot{E_{k}}$ are given by:

\begin{equation}\label{eq:4}
\begin{split}
\dot{M}\simeq \Omega \times 8600\ M_{\odot}\ yr^{-1}
\end{split}
\end{equation}

\begin{equation}\label{eq:5}
\begin{split}
\dot{E}_{k}\simeq \Omega \times 2.5 \times 10^{47}\ erg\ s^{-1}
\end{split}
\end{equation}

Multiple surveys have been done to find that \civ\ BALs appear in approximately 20\% of all quasars \cite[e.g.,][]{Hewett03,Dai08,Dai12,Gibson09,Allen11}. The outflows reported here are consistent with \civ\ BALs since we observed absorption troughs from \oiv\ \ly 787.71 which has a similar IP to \civ\ \ly 1548.19. Moreover, the populations of \civ\ and \oiv\ are similar over a broad range of \Uh\ \cite[see figure 12 in][]{Muzahid13}. Also, using the classification scheme in Paper I and the widths of their \neviii\ absorption trough widths in table \ref{tb:OutflowSystems}, all of the outflows here are BALs except for S3 which is a mini-BAL. Therefore, $\Omega$ of these outflows should be similar to that assumed for \civ\ BAL outflows. Thus, we adopt an $\Omega$ = 0.2 and this lead to $\dot{M}$ = 4300 $M_{\odot}\ yr^{-1}$ and $\dot{E}_{k}$ = 10$^{46.7}$ erg $s^{-1}$ for outflow S1b. These are the largest $\dot{M}$ and $\dot{E}_{k}$ reported for quasar outflows to date.

The former records are $\dot{M}$ $\sim$ 3500 $M_{\odot}$ yr$^{-1}$\cite[][]{Maiolino12} and $\dot{E_{k}}$ $\sim$ 10$^{45.9}$ erg s$^{-1}$ \cite[][]{Chamberlain15b}. Moreover, S1b has a $\dot{E}_{k}$ $\sim$ 20\% of its L$_{\text{Edd}}$. With this high $\Gamma_{\text{Edd}}$, S1b can be the dominate sources of energy for AGN feedback in the host galaxy of SDSS J1042+1646 \cite[][]{Scannapieco04}.

Similarly, for component 1a, this yields $R$ $=$ 840$^{+500}_{-300}$ pc, $\dot{M}$ $=$ 2800$^{+200}_{-800}$ $M_{\odot}$ yr$^{-1}$ and log($\dot{E_{k}}$) $=$ 46.4$^{+0.1}_{-0.1}$ erg s$^{-1}$. Combining S1a and S1b, outflow S1 has an $\dot{E_{k}}$ close to 10$^{47.0}$ erg s$^{-1}$ and $\Gamma_{\text{Edd}}$ close to 30\%. 

%These incredibly high $\dot{E_{k}}$ and $\Gamma_{\text{Edd}}$ suggest that S1 can provide more than enough energy needed for AGN feedback.

%For S2, $R$ = 17$^{+10}_{-10}$ pc, $\dot{M}$ $=$ 90$^{+20}_{-30}$ $M_{\odot}$ yr$^{-1}$, log($\dot{E_{k}}$) $=$ 45.2$^{+0.1}_{-0.2}$ erg s$^{-1}$, and $\Gamma_{\text{Edd}}$ $\sim$ 0.4\%.

%For S3, since \ne\ is undetermined, we do not have information to constrain $R$ and the energetics.

%Significant AGN feedback effects typically requires that the energy input, i.e, $\dot{E_{k}}$, is roughly 0.5 -- 5 percent of the L$_{Edd}$ of the quasar \cite[][respectively]{Hopkins10, Scannapieco04}. We derive the bolometirc luminosity (L$_{bol}$) for SDSS J1042+1646 as 10$^{47.17}$, which is around 40\% of L$_{Edd}$. Two of the outflows from SDSS J1042+1646 are very energectic with log($\dot{E_{k}}$) close to 10$^{47.2}$ erg s$^{-1}$. Therefore, with up to log($\dot{E_{k}}$) $\sim$ 38\% L$_{Edd}$, these two outflows can contribute significantly to the theoretically invoked AGN feedback effects.

%\subsection{\ov*\ and the Very High \ne\ Outflow in \XUV}

\subsection{Effects of the Super Solar Metallicity}
%For components 1a and 1b in outflow S1, the current HP phase has difficulty satisfying both N$_{ion}$ measurement of \siv\ and the N$_{ion}$ upper limit of \oiii\ (see figure \ref{fig:comp1}). A physically plausible explanation is to invoke a super solar metallicity (SSM). Moreover, the relative abundance of sulfur to oxygen can increase with increasing metallicity \cite[][]{Ballero08}. Therefore, SSM can shift the \siv\ curve closer to the \oiii\ curve and solve the discrepancy mentioned above.
%Another issue in outflow components 1a and 1b is that their current photoionization solutions produce $\sim$ 20 times more N(\hi) than the AOD values observed. This issue can also be fixed by invoking the SSM. However, in order to not introduce another degree of freedom, we only report the solutions assuming solar metallicity in this paper. 

AGN outflows can exhibit super solar metallicity (SSM) \cite[e.g.,][]{Arav07, Gabel06}. Paper V discusses the quantitative effects of SSM on the inferred parameters for outflows observed in the EUV500. Unlike the case shown in Paper V, the photoionization solutions of S1, S2, and S3 in SDSS J1042+1646 are consistent with solar metallicity. To estimate the effects of possible SSM on S1, S2, and S3, we follow approach from Paper V and compute the analysis results for two metallicity values: solar (Z$_{\odot}$) and 4.7 times solar (4.7Z$_{\odot}$). For outflow system 1 and 2, the results based on 4.7Z$_{\odot}$ leads to a decrease of distance by up to 30\% and decreases of $\dot{M}$ and $\dot{E}_{k}$ by up to a factor of 6. However, the goodness of the fit (determined by their $\chi^2$) for all models is comparable, so the fits cannot provide a metallicity constraint.

\section{Summary}

In this paper, we presented the analysis of HST/COS spectra for the quasar outflows seen in SDSS J1042+1646 in the \XUV\ (500 -- 1050\angstrom\ rest-frame) region. The results are summarized as follows:\\

1. A total of five outflow systems are identified (S1 -- S4, where S1 has two components: S1a and S1b). From these outflows, we observed absorption troughs from both high-ionization species, e.g., \oiv, and \siv, and very high-ionization species, e.g., \arviii, \neviii, \naix, and \mgx. Four out of the five outflows are defined as BAL outflows while the other (S3) is a mini-BAL outflow (see section \ref{sec:defineV}).

2. We developed the Synthetic Spectral Simulation method to efficiently fit the multitude of observed troughs with a three-dimensional model grid in the parameter space of \Uh, \Nh, and \ne. This method is especially powerful when handling blended absorption troughs from many EUV500 species ($>$70), and intervening absorption troughs (see section \ref{sec:SSS}). 

3. The appearance of the very high-ionization species necessitates at least two-ionization phase solutions for the observed outflows. Combining with previous studies, we suggest that all HiBALs have very high-ionization phases, which are almost exclusively accessible with EUV500 observations from HST (see section \ref{text:AllComps}).

4. With determining \ne\ from density sensitive transitions, we found the distances and energetics for 3 out of the 5 outflows. Outflow system 1 has $\dot{E_{k}}$ close to 10$^{47.0}$ erg s$^{-1}$, which is the most energetic outflow observed in quasars to date. Moreover, this leads to an $\dot{\text{E}_{k}}$/L$_{\text{Edd}}$ close to 30\%, which makes it a good candidate for being the agent of quasar-mode AGN feedback (see section \ref{sec:energy}). 

5. When we take into account the attenuation of ionizing flux by interior outflow systems, we find a decrease of $\sim$ 0.1 dex in R, $\dot{M}$ and $\dot{E_{k}}$ (see section \ref{sec:shading}).

\acknowledgments
X.X., N.A., and T.M acknowledge support from NSF grant AST 1413319, as well
as NASA STScI grants GO 14777, 14242, 14054, and 14176, and NASA ADAP 48020.

Based on observations made with the NASA/ESA \textit{Hubble Space Telescope}, obtained from the data archive at the Space Telescope Science Institute. STScI is operated by the Association of Universities for Research in Astronomy, Inc. under NASA contract NAS5-26555.

CHIANTI is a collaborative project involving George Mason University (USA), the University of Michigan (USA) and the University of Cambridge (UK).
%\bibliographystyle{emulateapj}
%\bibliographystyle{mn2e}
%\bibliography{astro}{}

%\appendix
%Here is the appendix section. 

\bibliography{apj-jour,dsr-refs}

\end{document}